\documentclass[11pt,a4paper]{article}

\usepackage[a4paper,margin=1.15in]{geometry}
\usepackage{xltabular}
\usepackage[T1]{fontenc}
\usepackage[utf8]{inputenc}
\usepackage{tocbibind}
\usepackage{url}
\usepackage{apacite}
\usepackage{graphicx}
\usepackage{comment}
\usepackage{float}
\usepackage[font=small,labelfont=bf,labelsep=period]{caption}
\usepackage{makecell}
\usepackage{mathtools}
\usepackage{commath}
\usepackage{booktabs}
\usepackage{multirow}
\usepackage{adjustbox}
\usepackage{enumitem}
\usepackage{bbold}
\usepackage{xcolor}
\usepackage{amssymb}
\usepackage{amsmath}
\usepackage{mathtools}
\usepackage{dutchcal}
\usepackage{tabularx}
\usepackage{threeparttable}
\usepackage{pdflscape}
\usepackage{subcaption}
\usepackage[bottom]{footmisc}
\usepackage{helvet} 
\fontfamily{phv}\selectfont
\usepackage[title]{appendix}
\usepackage{rotating}

\usepackage{placeins}
\let\Oldsection\section
\renewcommand{\section}{\FloatBarrier\Oldsection}

\RequirePackage{fix-cm} 
\DeclareMathSizes{12}{12}{10}{10}
\usepackage[doublespacing]{setspace}

\usepackage[ruled, vlined, linesnumbered]{algorithm2e}
\usepackage{etoolbox}
\AtBeginEnvironment{algorithm}{\setstretch{1.1}}
\SetCommentSty{\footnotesize\texttt}

\newboolean{showcomments}
\setboolean{showcomments}{true}
\ifthenelse{\boolean{showcomments}}
{ \newcommand{\mynote}[3]{
    \fbox{\bfseries\sffamily\scriptsize#1}
    {\small$\blacktriangleright$\textsf{\textit{\color{#3}{#2}}}$\blacktriangleleft$}}}
{ \newcommand{\mynote}[3]{}}

\newcounter{appendixtable}[section]

\newcounter{appendixfigure}[section]

\usepackage{chngcntr}
\counterwithin{table}{section}
\counterwithin{figure}{section}

\title{\textbf{Teamwork and Spillover Effects in Performance Evaluations}\thanks{We thank Riccardo di Francesco, Daniel Goller and Tommy Krieger for valuable comments on a previous draft of this paper.}}

\author{
	Enzo Brox \\ University of St.Gallen\thanks{University of St.Gallen, Swiss Institute for Empirical Economic Research, Switzerland; enzo.brox@unisg.ch.}
	\and
	Michael Lechner \\ University of St.Gallen\thanks{University of St.Gallen, Swiss Institute for Empirical Economic Research, Switzerland; michael.lechner@unisg.ch. Michael Lechner is also affiliated with CEPR, London, CESIfo, Munich, IAB, Nuremberg, IZA, Bonn, and RWI, Essen.} }

\date{\today}

\begin{document}

\clearpage
\maketitle
\thispagestyle{empty}
\begin{abstract}
 \onehalfspacing
 \noindent 
  
This article shows how coworker performance affects individual performance evaluation in a teamwork setting at the workplace. We use high-quality data on football matches to measure an important component of individual performance -- shooting performance -- isolated from collaborative effects. Employing causal machine learning methods, we address the assortative matching of workers and estimate both average and heterogeneous effects. There is substantial evidence for spillover effects in performance evaluations. Coworker shooting performance, meaningfully impacts both, manager decisions and third-party expert evaluations of individual performance. Our results underscore the significant role coworkers play in shaping career advancements and highlight a complementary channel, to productivity gains and learning effects, how coworkers impact career advancement. We characterize the groups of workers that are most and least affected by spillover effects and show that spillover effects are reference point dependent. While positive deviations from a reference point create positive spillover effects, negative deviations are not harmful for coworkers. 

\vspace{6pt} 
		\thispagestyle{empty} \noindent \textbf{Keywords}: \textsc{Teams, Peer effects, Performance evaluation, Causal machine learning} \newline
\end{abstract}

\newpage
\doublespacing
\pagenumbering{arabic}

\textit{"Great things in business are never done by one person. They're done by a team of people.”} 
\begin{flushright}
	(Steve Jobs, CEO, Apple Inc.) 
\end{flushright} 

\textit{"The nicest thing about teamwork is that you always have others on your side.”} 
\begin{flushright}
	(Margaret Carty, CEO, Maryland Library Association) 
\end{flushright} 

\section{Introduction}  \label{sec: Introduction}

Organizations increasingly embrace collaborative and team-based approaches to enhance productivity and promote innovation \cite{lazear2007, deming2017}. However, performance evaluations, which significantly impact workers' compensation and career advancement, typically take place on the individual level \cite{prendergast1999, dejanvry2023, frederiksen2020}. This challenges conventional management practices, as it requires rewarding and evaluating employees based on their contributions, even when these are not perfectly observed or separable from coworker influences \cite{jones2021}. Consequently, it creates dependencies between workers that are crucial to understand for organizations to accurately assess their employees and for researchers that seek to understand inequalities in career outcomes.



The objective of performance evaluations is to recognize valuable signals of a worker's performance. In a teamwork context, dependencies between coworkers make it difficult to obtain accurate information on individual contributions to team performance \cite{frederiksen2017}. This requires employers often to resort to subjective evaluations.\footnote{For a review of different approaches studying the use of subjective performance evaluations, see \citeA{kampkotter2016}. For another reference stressing the need of studying subjective performance evaluations due to its ubiquitous use, see \citeA{oyer2011}.} In this study, we investigate whether and how the subjective evaluation of the performance of a worker is affected by coworkers. We thereby test a complementary mechanism how coworkers exert an influence on the career outcomes of their peers beyond productivity and learning spillovers \cite{arcidiacono2017, jarosch2021}.\par

From a research perspective, pinpointing the impact of coworker performance on individual performance evaluations encounters several challenges. First, researchers do not have access to the complete array of information available to employers regarding the employees. Second, evaluating the individual performance of workers can be problematic, particularly within teamwork tasks where their contributions may be less discernible. Third, coworker performance is likely influenced by one's own performance, making it difficult to untangle the extent to which coworker performance should be attributed to one's own evaluation. Fourth, coworker assignment is not random and high-skilled workers are likely to be matched with high-skilled coworkers. Thus, the assortative matching of employers poses challenges to the causal interpretation.\par

This paper addresses the aforementioned challenges by utilizing a comprehensive dataset on performance in sport contests. We focus on soccer (football, \textit{German Bundesliga}), and exploit the use of advanced player tracking technologies to measure a specific aspect of players' productivity -- shooting performance. Linking information about the quality of a chance to score a goal, defined as a probability of a shot resulting in a goal (\textit{expected goals}), and the outcome of the shot allows assessing the shooting performance of a player and abstracting it from collaborative effects.\footnote{The idea behind this approach is similar to evaluating chess moves with a chess engine \cite{kunn2021}, where the pre move positioning (and post move positioning) of chess figures is used to evaluate the quality of chess moves.} We argue that, from the perspective of a player on the field, coworkers shooting performance can be considered as randomly assigned. 
We address the assortative matching of workers using detailed information on worker quality, match circumstances, manager characteristics, and methods from the causal machine learning literature that enable flexible estimation of average and heterogeneous causal effects. 




Although the shooting performance of coworkers is beyond the control of the individual worker, it significantly influences judgments and decisions related to players’ performance. Coworker performance has a meaningful impact on manager decisions regarding whether to field a player in the next match. Comparing the highest degree of coworker performance to the lowest degree of coworker performance under comparable circumstances decreases the probability to be replaced in the starting line up on average by 13\% relative to the sample mean. The size of the effect is equivalent to one-third of the effect of the own performance in the same task. While managers have strategic interest to field the most successful team in the upcoming match, also expert journalists' evaluations of a players' performance are substantially affected by coworker performance. Again, comparing the highest degree of coworker performance to the lowest degree of coworker performance leads to an almost 11\% increase in the own evaluation. We present evidence against alternative interpretations for our results and conduct various robustness checks to support our identification strategy. \par

We further investigate whether good and bad performances cause asymmetric spillover effects. A large literature documents that behavior in various domains is reference point dependent (e.g., \citeA{gneezy1997, pope2011}). We use the number of expected goals coworkers are exposed to as a natural reference point, and investigating whether positive and negative deviations from this reference point cause asymmetric spillover effects. There is substantial evidence for asymmetric effects. Positive deviations have a substantial positive impact on coworkers. Negative deviations, however, do not generate a negative impact on coworkers. For journalists' evaluations, large negative deviations from the reference point are even less detrimental than small negative deviations.

Despite documenting spillover effects in performance evaluations, it is important to understand which type of workers are most and least affected by spillover effects and to describe circumstances that trigger spillover effects. We exploit recent advances in causal machine learning to investigate heterogeneous treatment effects.

The first important observation is that the vast majority of players are affected by spillover effects. In addition, for manager decisions, player strength and tenure are negatively correlated with the size of the spillover effect, suggesting that new and less skilled players are affected most by coworker performance. This is in line with the idea that managers have less trust or less information about these players. We also find stronger spillover effects in lower quality teams, where good performance and success are less frequently observed. However, there is no evidence that the responsibility for a similar task is correlated with the effect size.\footnote{Shooting performance is primarily important for offensive players who's main task is to score goals, while a defensive player's main task is to prevent the opposing team from scoring goals.} 

For journalists' evaluations, we find that individual characteristics like player skill-level or club-tenure are not related to the effect size. However, spillover effects are highest for players who are responsible for a similar task (offensive players). Furthermore, offensive players are not only most affected by spillover effects, but theirs performances also exhibit the highest spillover effects on other players. Results from a policy simulation exercise and a k-means clustering approach to characterize the most and least affected groups of players support our findings.


In the last part, we examine potential long-term effects. While fielding decisions by managers already have career implications, as players receive more playing time, season-level aggregates of our treatment are also associated with elevated player evaluations over the following two years. Hence, spillover effects from coworker performance in a teamwork setting possess the potential to impact worker careers. This underscores an additional channel through which coworkers shape the career outcomes of their peers beyond their well documented effects on productivity \cite{mas2009, gould2009} and skill acquisition \cite{amodio2023b, jarosch2021}.\par 

Since the findings of this paper are obtained from an industry where every action of a worker is observable, we argue that spillover effects from coworker performance are even more likely to occur in other types of organizations. In standard organizational settings, worker performance in a teamwork task is less observable compared to our setting. This increased difficulty in objectively rewarding individuals for their contributions to team performance amplifies the reliance on subjective judgments. To the best of our knowledge, this is the first paper studying the direct impact of coworker performance on individual performance evaluation in a teamwork setting and providing evidence for spillover effects in individual performance evaluations at the workplace.\par 

\subsection{Relation to literature}

Our study contributes to various strands of the literature. First, we add to the existing research on peer effects in the workplace. There are multiple potential factors that can result in productivity spillovers among coworkers. One such factor is the nature of the production technology itself, where coworkers impacting each other's marginal productivity \cite{gould2009, arcidiacono2017}. Additionally, observing coworkers' efforts motivates individuals to exert more effort themselves \cite{kandel1992, falk2006, cohen2023effort}.\footnote{The designs of pay structures, dismissal policies and input heterogeneity also generate externalities among workers. For example, relative performance evaluations and team-based assessments create interdependencies in coworkers' effort choices \cite{bandiera2005, mas2009, amodio2018}.} Beyond contemporaneous productivity spillover effects, several recent studies emphasize the additional benefits of peer learning and knowledge transfer among workers \cite{papay2020}. \citeA{nix2020} found that an increase in the average education level of peers positively impacts earnings in subsequent years. \citeA{akcigit2018} demonstrated that exposure to accomplished patent holders increases the likelihood of patenting and improves the quality of patents, underscoring the role of knowledge diffusion through peers. \citeA{herkenhoff2018} explored learning with production complementarities, revealing how workers' skills can be enhanced through peer interactions.\footnote{Outside workplace settings, a substantial body of literature in education economics delves into contemporaneous and lasting peer effects. For a review, refer to \citeA{sacerdote2014} and \citeA{bramoulle2020}.} Additionally, \citeA{jarosch2021} documented a robust association between having highly paid coworkers and future wage growth, highlighting spillover effects in terms of career advancement. 

Our study contributes to the existing literature by revealing an additional channel through which workers shape the career outcomes of their peers. In a team-based setting, we showcase a significant impact of coworkers on the performance evaluations of their peers, extending beyond productivity spillovers. Given the pivotal role of performance evaluations in compensation and career advancement, our findings underscore the importance of coworkers, especially in teamwork contexts. While our results do not negate the existence of learning from coworkers, they highlight an additional and potentially widespread mechanism through which workers influence their peers' career outcomes in modern organizations. These findings align with the reduced-form evidence presented in \citeA{jarosch2021} using German administrative labor market data and speak to a broader literature studying how heterogeneity in workplace conditions affects career outcomes \cite{arellano2024, vonWachter2020}. \par

Second, our findings contribute to the literature on performance evaluations, which are pivotal in organizations for determining compensation, promotion opportunities, and providing employee feedback \cite{dejanvry2023, frederiksen2020}. The existing literature has explored various dimensions of performance evaluations. First, research on performance evaluation design has examined different approaches, such as relative performance evaluations, objective metrics, and multidimensional assessments \cite{lazear2007, baker2003, cascio2008, manthei2019}. Second, studies have shown that performance feedback influences employee motivation, job satisfaction, and subsequent performance \cite{ilgen2005, kampkotter2018}. Third, biases in performance evaluations have been extensively studied \cite{ilgen1983, gauriot2019, principe2022}.\footnote{A related literature studies whether women and men are differentially recognized for performance \cite{sarsons2021, card2019, card2022}.} Several recent studies use data from the sports industry to study biases in performance evaluations, due to observability of individual performance. \citeA{principe2022} uses data from the top Italian soccer league to show that black players receive worse evaluations compared to non-black players, despite no differences in performance. \citeA{gauriot2019} and \citeA{lefgren2015} provide evidence for an outcome bias using data from professional soccer and basketball. They exploit quasi-experimental settings where luck, rather than effort and performance, determines the outcome, affecting performance evaluations. This is argued to violate the informativeness principle \cite{holmstroem1979}, which states that only valuable signals for performance should be rewarded.\footnote{There is also a substantial psychological literature on the outcome bias. Following \citeA{baron1988}, several studies have suggested that individuals tend to give too much importance to information about the outcome when trying to assess the quality of decisions made by an agent (e.g. \citeA{alicke1994, gino2010}).} We contribute to this literature by studying spillover effects in performance evaluations in a teamwork setting and by providing evidence that coworkers present another source of bias in individual performance evaluations. \par

Third, our findings contribute to a broader literature on teamwork spanning various disciplines, including economics, business, organizational behavior, and psychology. The economic literature extensively examines the benefits of teamwork in terms of enhancing productivity and fostering innovation \cite{smith2000, lazear1999}, focusing on the division of labor, specialization, and coordination mechanisms within teams \cite{akerlof2005, weidmann2021, marx2021, brox2022}. From a business perspective, research explores how effective teamwork can lead to improved organizational performance, customer satisfaction, and competitive advantage \cite{katzenbach2015, hackman2005}. The psychological literature examines individual and team-level factors impacting team effectiveness, including team cohesion, trust, conflict management, and motivation \cite{mathieu2017, west2012}.\par 

This paper contributes to this literature by examining spillover effects in teams. Given the non-trivial nature of this task, the empirical literature on peer effects in teamwork settings is relatively scarce \cite{cornelissen2017}. \citeA{weidmann2021} employ an experimental setup to study spillover effects in teamwork, demonstrating that individuals with higher socio-emotional skills have the potential to significantly enhance team performance. \citeA{arcidiacono2017} use data from the North American Basketball League (NBA) to illustrate substantial differences among basketball players in their contributions to the performance of their peers, also arguing that players receive less credit for spillover effects than for own performance. \citeA{cohen2023effort} provide evidence of peer effects in effort provision using data from the professional Israeli soccer league. Additionally, \citeA{amodio2023b} reveal skill adoption patterns among coworkers using data from the North American Hockey League (NHL). Our contribution to the understanding of teamwork settings lies in revealing externalities arising from coworker performance.\par

The remainder of the paper is structured as follows. Section \ref{sec: Institutional Framework} outlines the key elements of the German \textit{Bundesliga}. Section \ref{sec: Data and Measuring Shooting Performance} introduces the datasets employed in our empirical analysis. Section \ref{sec:Empirical Strategy} details the empirical strategy and estimation procedures. Section \ref{sec: Results} presents the findings of our empirical analysis. In Section \ref{sec: Career effects}, we concentrate on long-run effects. Finally, Section \ref{sec: Conclusion} concludes.

\section{Institutional Framework}   \label{sec: Institutional Framework}
 
To explore the impact of coworkers on individual performance evaluations, we utilize data from the \textit{Bundesliga}, the premier division of German male football. The \textit{Bundesliga} is widely regarded as one of the top-five football leagues in Europe and consists of 18 clubs. The league operates in a double round-robin format, with each club facing all other opponents twice in a season, totaling 34 matches per club evenly distributed between home and away games. Matches predominantly occur on weekends, with the first round typically spanning from August to December, followed by a brief winter break. The second round runs from January to May/June, with a change in the home field advantage between the two rounds.\par

Teamwork in soccer is fundamental to the success of a team. It involves coordinated efforts among players to achieve common objectives, such as creating chances to score goals and preventing the opposing team from scoring goals. Each match comprises two halves of 45 minutes each, and a team secures a victory by scoring more goals than its opponent.  The winning team earns three ranking points, while the losing team receives none. In the event of a draw, both teams are awarded one point. The final league position at the end of the season is determined by the total accumulated ranking points, influencing relegation decisions and participation in international competitions in the following season. Before each match, the team manager selects eleven starting players — a goalkeeper and ten field players (defender, midfielder, attacker) — from the available squad. Additionally, up to seven potential substitutes can be nominated, with the opportunity for up to three substitutions during the match. 

\par

\section{Data}   \label{sec: Data and Measuring Shooting Performance}

We utilize data spanning from the 2010-2011 season to the 2019-2020 season, with the primary data source being Opta, a private company specializing in data collection and distribution for various sports.\footnote{https://www.statsperform.com/opta/} Opta specializes in gathering in-play information for football matches, capturing details such as passes, duels, and shots on goal. With advanced tracking technology, they record spatial coordinates, providing a comprehensive overview of player actions during matches. The dataset includes the names of the involved players in every event of the match, the time of each action, and the corresponding spatial coordinates.\footnote{Event data quantifies what happens on the pitch in relation to individual player actions. For example, each time a player passes the ball, tackles an opponent, shoots on goal, or saves a shot is one event.}\par

The analysis relies on two crucial features of this dataset. First, it provides comprehensive information on every shot taken, encompassing the player's name, the outcome (whether it resulted in a goal or not), and a quantitative assessment of the quality of the chance from which the shot resulted (further details below in Section \ref{subsec: Identification} and in Appendix \ref{subsec: Appendix B}). We exploit this information to measure a player's shooting performance. Second, the dataset offers detailed performance measures at the individual level. Through Opta's advanced tracking technology, player tracking is conducted throughout the match, resulting in over 100 categories of performance metrics at the individual level. This level of granularity provides precise insights into the actions of each player. In addition to standard measures like goals scored, it encompasses less observable performance indicators such as attempted dribbles and successful crosses. A comprehensive description of these performance measures is relegated to Appendix A (Table \ref{table_metrics1} and Table \ref{table_metrics2}). We also gather information on each team's line-up and substitutions from this dataset.\par

We enhance this dataset by incorporating additional information from various sources. First, we gather background details on the players, including their date of birth, date of hiring by a Bundesliga club, playing position, preferred foot, nationality, and physical attributes. Furthermore, the data is augmented with two time-varying expert-based indicators of player quality. One measure is derived from \textit{www.transfermarkt.com}, a reputable platform for tracking player market values. The other measure originates from the popular video game FIFA, developed by Electronic Arts. Each player is assigned a playing strength rating ranging from 0 to 100. This measure offers notable advantages compared to market values. The game is widely played globally and strives for realistic player evaluations, motivating game developers to ensure accuracy. Additionally, the measure is age-independent. Finally, experts evaluate players across various game-relevant attributes that capture different aspects of a player's skills. These include shooting skills, passing skills, running skills, dribbling skills, physical skills, and defensive skills \cite{brox2022}. An example of a player's evaluation is illustrated in Figure \ref{fig: fifa_eval} in Appendix A. To avoid endogeneity issues, we utilize quality measures updated before the start of each half-season.\par 

Second, we incorporate player ratings from \textit{Kicker}, a popular sports newspaper in Germany. After each match, since 1963, independent experts assign ratings to players on a scale of 1 to 6 in 0.5 increments, with 1 being the best grade and 6 being the worst grade. The grades appear in the online and print version of the newspaper after each match day. For each match, there are two experts assigned, and the observed grade is the joint evaluation of both experts. All players who played for at least 30 minutes in a match receive a rating. For an example of such an evaluation, see Figure \ref{fig: kicker} in Appendix A.\par

Third, we enrich our dataset with match-related information to characterize match circumstances. This includes for example whether a club has previously participated in international competitions or is scheduled to participate in an international match following the current one. Furthermore, we calculate the number of days between each team's previous and next match, capture whether a match is played on the home ground, the ranking position of the participating teams, whether a team has been relegated in the season before, and details regarding the manager's experience. For a comprehensive list, please refer to Table \ref{table_desc_stat2} in Appendix A.\par

\section{Empirical Strategy}    \label{sec:Empirical Strategy}

\subsection{Identification}      \label{subsec: Identification}

Identifying the effect of coworker performance on individual performance evaluation is challenging for at least two reasons. First, in a teamwork setting, coworker performance is likely to depend on individual performance. This makes it difficult to isolate coworker performance from individual performance, which naturally influences performance evaluations. Second, coworker performance is likely to depend on coworker quality. As coworkers are not randomly assigned, this introduces the potential for selection bias when estimating the effect of coworker performance on individual performance evaluations. The remainder of this section describes how we address these challenges.

To address the former challenge, we leverage the richness of our data, enabling us to measure the performance of workers in one specific dimension isolated from coworker influences. From the point of view of the individual worker, coworker shooting performance is as good as randomly assigned. For simplicity, consider player A playing alongside player B. In two subsequent matches, player B faces a chance to score a goal from a similar position. In the first match, player B scores the goal, while in the second match, he fails to score.\footnote{Failing to score can be the result of multiple issues. Player B may either perform worse, or be less lucky, or the opponent team (goalkeeper) shows a better performance. However, none of these issues is influenced by player A.} The identifying assumption rests on the fact that we can adequately control for the performance of player A in creating the scoring opportunity for player B. This allows us to control for factors that jointly determine shooting performance, as well as player A's performance evaluation (details are discussed in Section \ref{subsec: Measuring Shooting Performance} below).

Although shooting performance is not influenced by coworkers, the assignment of workers to coworkers is not random. Assortative matching between workers and coworkers results in a positive association between coworker quality/performance and performance evaluation. We address this issue using state-of-the-art methods from causal machine learning and detailed information on worker quality, teams and match circumstances. In particular, we control for the quality of workers in all our specifications in a flexible way and also include controls that describe the circumstances of the match, such as the overall quality of the teams, their current form, the regeneration time of each team before and after a match, and whether a team faces important matches before and after the match.\footnote{Note that since we control in a detailed way for the quality of the worker, we eliminate concerns regarding worker quality jointly affecting coworker performance and performance evaluation.}

A remaining concern regarding the argument that coworker performance directly affects individual performance evaluation is that coworker quality, even after addressing assortative matching of workers, may affect individual performance and thus individual performance evaluations. We consider this unlikely since shooting performance is likely to be determined by shooting ability, which is usually not the dominant skill/ability that creates the largest spillover effects. However, shooting ability may be correlated with other skills that are more prone to create coworker spillover effects in productivity. We address this concern in the robustness checks in two ways. First, by adding detailed coworker quality control measures. Second, we exploit the richness of the data and include a large set of individual performance metrics. This should eliminate the concern that productivity spillovers resulting from day-to-day variation in coworker quality are driving our result (Table \ref{table_estimation_results_main2}).

\subsection{Shooting Performance}  \label{subsec: Measuring Shooting Performance}

Measuring individual performance within a team-based task presents inherent challenges due to the complex nature of performance observation and its reliance on multiple contributors. Take, for instance, the example of a soccer player's pass. Whether player A is able to play a successful pass hinges not only on the passer's skill but also to a significant extent on the recipient's positioning and decision-making, as well as on the quality of the previous pass before player A received the ball. In this study, we leverage recent advancements in player monitoring within the sports industry and exploit a method that enables the separation of players shooting performance from collaborative effects.
To do so, we leverage a widely used metric in modern sports analytics called 'expected goals' (xG). 
\par

The xG metric is extensively used in predictive modeling, player, and team analysis. It quantifies the quality of a scoring opportunity at the moment when a shot is taken, by employing machine learning to estimate the probability of a shot resulting in a goal. These predictions are derived from a training data set containing information from thousands of shots. Importantly, xG probabilities are not influenced by the player's skill level; instead, they consider factors such as the shot's location (distance and angle to the goal), the positioning of other players of both teams, the goalkeeper's position, the type of assist, the shot type, and the flow of the game in the second when a shot is taken.\footnote{For a more comprehensive explanation of how expected goals are calculated, refer to Appendix \ref{subsec: Appendix B}.} \par

\begin{figure}[h!]
    \caption{Expected Goals}
    \begin{subfigure}{0.45\textwidth}
        \centering
        \includegraphics[width=\textwidth]{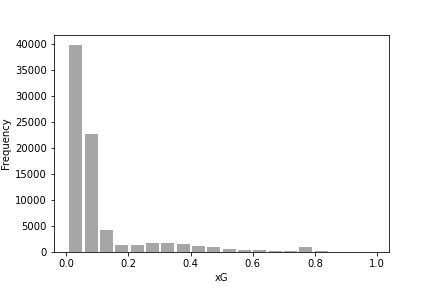}
        \caption{Shots}
        \label{fig:expected_goals_shots}
    \end{subfigure}
    \hfill
    \begin{subfigure}{0.45\textwidth}
        \centering
        \includegraphics[width=\textwidth]{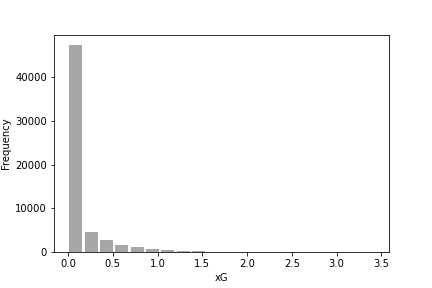}
        \caption{Individual}
        \label{fig:expected_goals_ind}
    \end{subfigure}
    \hfill
    \begin{center}
    \begin{subfigure}{0.45\textwidth}
        \centering
        \includegraphics[width=\textwidth]{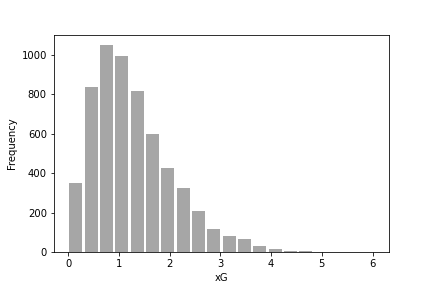}
        \caption{Team}
        \label{fig:expected_goals_team}
    \end{subfigure}
    \end{center}
    \footnotesize 
    \textbf{Notes:} This figure shows histograms of expected goals at different aggregation levels. Panel (a) displays the distribution of expected goal values at the shot level. Panel (b) aggregates expected goals at the individual-match level, while Panel (c) aggregates expected goals at the team-match level. 
    \label{fig:expected_goals}
\end{figure}

Figure \ref{fig:expected_goals} provides information on the xG metric. Figure \ref{fig:expected_goals_shots} shows the distribution of xG values over all shots in our sample. xG values range between zero and one, with the majority of values close to zero. The small spike between 0.7 and 0.8 constitutes penalties that have a fixed xG value. Figure \ref{fig:expected_goals_ind} displays the distribution of aggregated xGs per individual in a match, ranging from 0 (if a player does not face a chance to score) to around 3.4. Approximately 40\% of individual-match observations do not involve a chance to score.\footnote{In Appendix B, Figure \ref{fig: expected goals_by_pos} shows the distribution of xG by position. Reassuringly, strikers face the highest xG values, while defenders face the lowest xG values.} In Figure \ref{fig:expected_goals_team}, we show aggregated xG values by team in a match, ranging from 0 to around 6, with teams expected to score slightly more than 1 goal on average. \par


We utilize the assessment of chance quality via xG to quantify the shooting performance of individual players. Specifically, we measure shooting performance by considering the number of goals a player scores under the condition of the chance quality he encounters. This allows to isolate an important component of individual performance, in a teamwork setting, from the influence of coworkers.\footnote{We conduct a further test to verify that our approach indeed measures shooting performance by correlating expert assessments of shooting ability and our approach to measure shooting performance. For details, see Appendix B Table \ref{tab:ability_corr}.}

\subsection{Variable definitions} \label{subsec: Variables}

\subsubsection{Outcome variables}   \label{subsubsec: Outcome variables}

We employ two performance evaluation metrics: manager decisions to field players in subsequent matches and expert evaluations provided by journalists. In this section, we provide a more detailed description of these outcome variables.

Team managers play a crucial role in making decisions about player selection throughout the season. They are tasked with assessing each player's performance at a particular moment and making decisions based on these evaluations when forming match lineups \cite{gauriot2019}. Existing research suggests that the primary objective of team managers is to maximize the probability of winning matches, rather than solely focusing on profits \cite{sloane1971, garcia2009}. Therefore, managers are motivated to select what they perceive as the best players for each match. Our analysis focuses on the fielding decisions for starting line-ups, aiming to determine whether a player's chances of staying in the starting line-up vary based on the shooting performance of their coworkers.\footnote{It is important to note that we lack information about the availability of players not in the lineup for the next match. While we assume this absence is not correlated with coworkers' shooting performance due to its quasi-random nature, the lack of control for this factor may introduce noise, potentially affecting the likelihood of observing an effect, if one exists.}

The second outcome variable are journalists' ratings published in the sports magazine \textit{Kicker}. These ratings are presented after each match and represent a subjective assessment of the players' performance. To ensure easier interpretation, we invert the scale, with our measure ranging from 1 (poor) to 6 (excellent). There are important differences between journalists' evaluations and manager decisions. First, unlike manager decisions, journalists' evaluations do not have immediate career consequences. Second, journalist do not have an incentive to field the most promising team for the next match. However, the evaluation of soccer players by journalists is subject to public debate  and forms the basis for an interactive online manager game. Therefore, journalists have incentives to evaluate as objectively as possible the individual performance of each player.\footnote{https://www.sueddeutsche.de/sport/fussball-bundesliga-noten-vielleicht-reicht-auch-mal-eine-fuenf-1.141370}

\subsubsection{Coworker Performance}    \label{subsubsec: Treatment}

The objective is to investigate the influence of coworkers' shooting performance on performance evaluation. The shooting performance of coworkers is measured as follows. We utilize data on all 79,876 shots taken over the ten seasons of German soccer and leverage information on the outcome of each shot, the quality of the chance (xG), and the player who took the shot. For each player $i$ and match $m$, we calculate the sum of goals scored by all his coworkers $j$ in the starting lineup: $GC_{im} = \sum_{j \neq i} \text{goals}$. This variable serves as the treatment variable in the main analysis.

To isolate the shooting performance of coworkers from the individual's own performance, a crucial control variable is the measure of the quality of chances that all coworkers $j$ of an individual $i$ faced. We measure the quality of chances that coworkers were exposed to in the same way as the number of goals scored by coworkers. For each player $i$, we calculate the sum of xG that his coworkers $j$ in match $m$ were exposed to: $xGC_{im} = \sum_{j \neq i} xG$. This control variable is used in all specifications.\footnote{We show an alternative specification in which we define coworker performance directly via the difference between $GC_{im}$ and $xGC_{im}$ in Section \ref{subsec: Nonlinear}.}

\subsubsection{Control variables}   \label{subsubsec: Control variables}

To approximate the identification prerequisites discussed in Section \ref{subsec: Identification}, a vector of pre-determined variables $X_i$ is chosen to control for assortative matching of coworkers. To control for assortative matching of workers into teams, a set of individual background characteristics is included in this vector. This set of variables includes various measures to control for worker quality and skills in various categories, as discussed in detail in Section \ref{sec: Data and Measuring Shooting Performance} above. It also includes measures of the time a worker is already in the club, their age, nationality, and physical composition. The overall strength of the team is controlled for with a pre-season measure for team strength that remains constant over the half-season. To control for sorting in particular matches, a detailed set of characteristics of the match circumstances is included in $X_i$. These include the season and half-season in which the match takes place, whether a match is played at home, the position of both teams in the league table, whether the team is engaged in international competitions, information on whether international matches have been taking place before or after a match, the number of days between the last and the current match, and the number of days between the current and the next match to control for potential strategic fielding decisions. Since the manager takes the decision on which players to field, information on the time the manager is in the club and about his experience as a coach is also added. For the complete list of control variables used in the specification, refer to Table \ref{table_desc_stat2}.

To avoid biases due to using bad controls, we focus on control variables, which are predetermined before the observation period (match) \cite{angrist2009}. Nonetheless, to enhance the robustness of the findings, we conducted additional analyses by expanding the vector of controls to encompass factors such as individual performance metrics in the match.

\subsection{Sample}    \label{subsec: Sample}

In general, information is available for all players who participated in a match during the seasons 2010 to 2020. The sample is restricted in several ways. First, it is limited to starting players. This simplifies the interpretation of the managerial decision on whether to field a player in the next match and allows interpreting the results as changes in the probability of staying in the starting line-up. Additionally, for almost all players, the necessary information is available, as players have to play at least 30 minutes to receive a grade. This ensures a comparable sample for both outcome variables. Second, goalkeepers are excluded from the analysis since they perform a very different task, are rarely replaced, and the quality measures cannot characterize them in the same way as other players. Third, all matches from the last match day of a season are dropped because the starting line-up in the next match is not observed. The final sample consists of 58,968 player-match observations.

\subsection{Estimation}     \label{subsec: Estimation}

To assess the impact of coworker performance on performance evaluations, we leverage advancements in the field of causal machine learning. The integration of techniques from machine and statistical learning (e.g., \citeA{athey2019machine}) with micro-econometric methodologies that aim to identify causal effects (e.g., \citeA{imbens2009}), offers distinct advantages over traditional estimators. 

We utilize the Modified Causal Forest (\textit{mcf}) estimator, introduced by \citeA{lechner2018} and further investigated by \citeA{lechner2024}.\footnote{In a Monte Carlo analysis they demonstrate the good performance of the (\textit{mcf}) compared to other popular choices such as Double/debiased machine learning \cite{chernozhukov2018double} or the Generalized Random Forest \cite{athey2019generalized}.} Compared to other causal machine learning estimators (e.g. \citeA{athey2019generalized}), an important advantage of the \textit{mcf} is that it provides a framework to estimate all causal parameters of interest in one unified estimation and a straightforward approach to perform unified inference across all aggregation levels.\footnote{We show results for Average Treatment Effect (ATE), Group Average Treatment Effect (GATE), and Individualized Average Treatment Effect (IATE). The IATEs measure the treatment effect for units with features $x$ and present the causal parameters at the finest aggregation level. For the GATEs we preselect variables according to research and practitioners interest and show effect heterogeneity with regard to the chosen parameters.} Furthermore, the \textit{mcf} allows for applications in empirical settings involving multiple discrete treatment variables, which is especially useful for the empirical application in our study.\footnote{For a comprehensive understanding of the estimator's technical aspects, the interested reader is directed to \citeA{lechner2018} and \citeA{lechner2024}. Additionally, for more in-depth insights into the implementation of the Python package (\textit{mcf}) available freely by PyPI, readers find detailed information in \citeA{bodory2022}.} We also test the robustness of the results using a traditional linear regression framework and an alternative machine learning estimator \cite{kennedy2017} in Section \ref{subsec: Main results} and Section \ref{subsec: Nonlinear}.

For estimation purposes the treatment is restricted to the categories, zero, one, two or more than two goals scored by coworkers, due to the lack of common support on the higher levels of the treatment variable. Common support figures for our preferred specification are subsumed in Figure \ref{fig: common_support}. To show that results are not driven by extreme values, we perform a robustness check and cut the sample and use only observation in which coworkers score zero, one or two goals (Table \ref{table_estimation_results_margin}).\footnote{Only around 10\% of observations have coworkers scoring more than two goals in a game.} In general, specifications differ by the set of covariates included to address assortative matching and to rule out alternative interpretations.

\subsection{Descriptive Statistics} \label{subsec: Descriptive evidence}

\begin{figure}[h!]

      \caption{Association of coworker performance and individual performance evaluation}
      \label{fig_corr_evidenz}
    \begin{subfigure}{.5\textwidth}
      \centering
      \includegraphics[width=\textwidth]{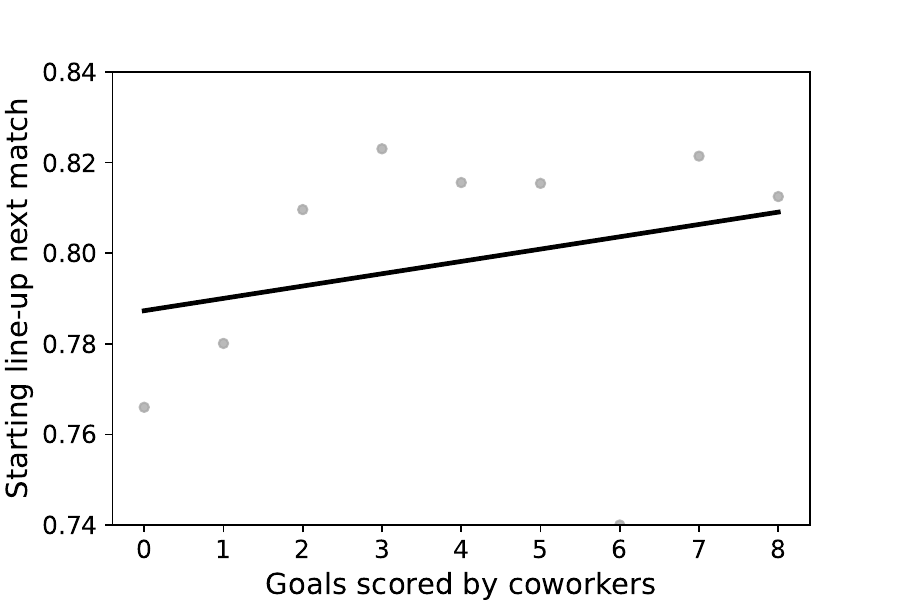}
      \caption{Manager}
    \end{subfigure}
    \hfill
        \begin{subfigure}{.5\textwidth}
      \centering
      \includegraphics[width=\textwidth]{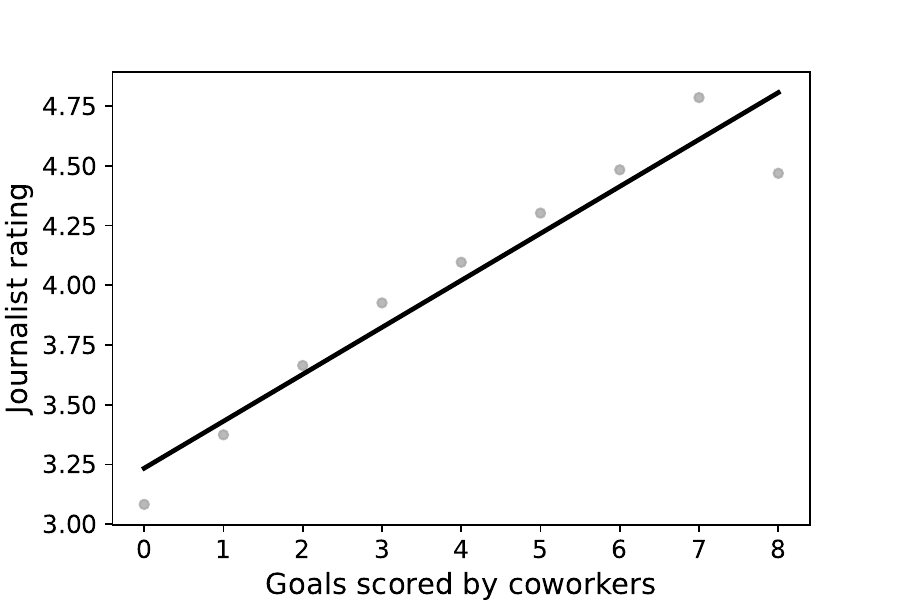}
      \caption{Expert}
    \end{subfigure}   
	\footnotesize \textbf{Notes:} This figure shows the correlation of coworker performance with performance evaluation for both outcome variables. The y-axis reports the mean values of the outcome variable for each level of the treatment variable.
\end{figure}

Figure \ref{fig_corr_evidenz} shows how shooting performance of coworkers correlates with individual performance evaluations. For this purpose, we plot the number of goals scored by coworkers against the own performance evaluation of a player for his expert evaluation in the newspaper and for the manager decision. We find a substantial positive correlation between coworker shooting performance and the own performance evaluations for both outcome variables. However, this result might largely depend on assortative matching of workers and does not account for the fact that goals are a product of team performance.

\begin{table}[H]
    \centering
    \footnotesize
    \begin{threeparttable}
    \caption{Descriptive Statistics of selected variables by Treatment Status}
\begin{tabular}{lcccccc}
	\tabularnewline 
    \toprule
	& \multicolumn{6}{c}{Goals scored by coworkers}  \\  
     \midrule 
	& \multicolumn{2}{c}{0}             & \multicolumn{2}{c}{1}              & \multicolumn{2}{c}{2}   \\  
	\midrule
	\emph{Outcome Variables}\\ \\
	Starting line-up next match &   0.77 & (0.42) &   0.78 & (0.41) &   0.81 & (0.39) \\
 Journalist rating &   3.08 & (0.93) &   3.37 & (0.91) &   3.80 & (0.89) \\   
    \midrule
	
	\emph{Covariates}\\ \\

    \emph{Chance quality}\\
Expected goals coworker &       0.70 &       (0.47) &       1.09 &       (0.57) &       1.73 &        (0.79) \\ \\
\emph{Individual controls}\\
        Position &   2.82 &  (0.74) &   2.77 &  (0.72) &   2.70 &  (0.69) \\
     Time in club (years) &   2.31 &  (2.16) &   2.41 &  (2.23) &   2.60 &  (2.41) \\
              Age (years) &  26.17 &  (3.68) &  26.14 &  (3.69) &  26.06 &  (3.74) \\
           German &   0.44 &  (0.50) &   0.44 &  (0.50) &   0.43 &  (0.50) \\
            Market value (ln) &  15.06 &  (1.08) &  15.17 &  (1.11) &  15.40 &  (1.19) \\
          Player strength (Fifa) &  74.33 &  (4.80) &  74.84 &  (4.91) &  75.76 &  (5.36) \\ \\
    \emph{Team controls}\\
   Rank last season &  10.35 &   (4.99) &   9.64 &   (5.09) &   8.39 &   (5.31) \\
    Number of league matches (manager) & 119.15 & (136.82) & 121.68 & (138.14) & 135.33 & (148.32) \\
      Number of Top-5 matches (manager) & 137.80 & (158.35) & 145.77 & (168.44) & 174.75 & (196.65) \\
     Average market value squad (ln) &  14.87 &   (0.71) &  14.99 &   (0.74) &  15.24 &   (0.84) \\
        Average market value Top 11 (ln) &  14.54 &   (0.76) &  14.67 &   (0.79) &  14.93 &   (0.88) \\
   Average squad strength (Fifa) &  70.62 &   (2.38) &  71.06 &   (2.48) &  71.76 &   (2.77) \\ \\
    \emph{Match controls}\\
    Home team &   0.43 &   (0.50) &   0.50 &   (0.50) &   0.58 &   (0.49) \\
   Rank last season (opponent) &   8.94 &   (5.37) &   9.43 &   (5.17) &  10.21 &   (4.91) \\
    Number of league matches manager (opponent) & 130.44 & (148.82) & 126.20 & (139.25) & 116.84 & (132.95) \\
      Number of Top-5 matches manager (opponent)  & 166.94 & (194.94) & 152.52 & (171.14) & 133.47 & (151.50) \\
     Average market value quad ln (opponent) &  15.14 &   (0.83) &  15.03 &   (0.77) &  14.89 &   (0.70) \\
        Average market value Top 11 ln (opponent) &  14.83 &   (0.88) &  14.71 &   (0.82) &  14.56 &   (0.75) \\
	\midrule
	Observations   &  \multicolumn{2}{c}{20,721}          &  \multicolumn{2}{c}{19,953}         &  \multicolumn{2}{c}{18,294}\\  
	\midrule
        \end{tabular}
        \begin{tablenotes}
            \footnotesize
            \item \textbf{Notes:} Means and standard deviations of selected variables by treatment status. Standard deviation are reported in parentheses. For the full table, see Table \ref{table_desc_stat2} in Appendix A. Position: = 2; if the main position is defender; = 3 if the main position is midfielder; = 4 if the main position is striker.
        \end{tablenotes}
        \label{table_desc_stat1}
    \end{threeparttable}
\end{table}

Table \ref{table_desc_stat1} shows the means and standard deviations (in brackets) by treatment status for a number of selected variables used in the main specifications that justify these concerns. Workers receive better evaluations when coworkers score more goals. However, coworkers were exposed to more/better chances to score goals, reflecting a better team performance. We also observe that players of higher quality are exposed to more goals by coworkers and teams that have ranked better in the last season, as well as teams playing on home ground and against weaker opponents. We address these issues in Section \ref{sec: Results}.\par

\section{Results} \label{sec: Results}

In this section, we present the results of the empirical analysis. Section \ref{subsec: Main results} shows the main results for the effect of coworker performance on individual performance evaluation. In Section \ref{subsec: Nonlinear}, we explore whether spillover effects are influenced by reference points, distinguishing between coworker performance below and above expectations. Section \ref{subsec: Heterogeneity} investigates heterogeneous treatment effects for a set of important variables. We also use more data driven approaches, like an optimal policy simulation and k-mean clustering, to investigate which group of players is most and least affected by coworker performance. Finally, Section \ref{subsec: Tasksandcloseness} investigates whether spillover effects depend on the main responsibility of workers or whether the proximity to coworkers affects the size of spillover effects.

\subsection{Main Results} \label{subsec: Main results}

Table \ref{table_estimation_results_main} presents the main findings across several model specifications. Throughout all specifications, we control for the quality of chances coworkers face (xG) to isolate coworker shooting performance. Columns differ in the way we account for the assortative matching of workers. The first column incorporates a comprehensive set of player characteristics to account for assortative matching. Columns two and three progressively include additional control variables at the team level (e.g. team quality, manager experience) and match level (e.g. opponent, time of the season), respectively.\footnote{For a comprehensive list of covariates included in each specification, see Appendix A Table \ref{table_desc_stat2}} The reported coefficients should be interpreted as a comparison of being exposed to the lowest level of coworker performance compared to being exposed to the highest level of coworker performance (specifically, 2 or more goals vs. 0 goals scored by coworkers from comparable situations). 

\begin{table}[H]
    \centering
    \begin{threeparttable}
        \caption{Average effect of coworker performance on manager decisions}
        \begin{tabular}{lccc}
            \toprule
            & \multicolumn{3}{c}{Starting line-up next match} \\
            \cline{2-4} 
            \addlinespace
            & (1) & (2) & (3) \\  
            \midrule
            Coworker performance & 0.038$^{***}$ & 0.036$^{***}$ & 0.033$^{***}$\\   
            & (0.012) & (0.011) & (0.010) \\ \\ 
            Mean probability to stay in the line-up & 78\% & 78\% & 78\%\\   
            \midrule
            Individual controls & $ \surd $ & $ \surd $ & $ \surd $\\ 
            Team controls & & $ \surd $ & $ \surd $\\ 
            Match controls & & & $ \surd $ \\ 
            Observations & 58,968 & 58,968 & 58,968\\  
            \midrule
        \end{tabular}
        \begin{tablenotes}
            \footnotesize
            \item \textbf{Notes:} *, **, and *** represents statistical significance at the 10\%, 5\%, and 1\% level respectively. Standard errors are clustered at the match level and are presented in parentheses. For a full list of covariates included in each specification, see Table \ref{table_desc_stat2}. All effects are population averages (ATE). 
        \end{tablenotes}
        \label{table_estimation_results_main}
    \end{threeparttable}
\end{table}

The results reveal a noteworthy and statistically significant positive effect of coworker shooting performance on worker performance evaluation. We find that when coworkers score two goals or more compared to coworkers scoring zero goals, from similar chances, there is an increase of over 3 percentage points in the likelihood for a worker to remain in the starting line-up. This effect is both statistically significant and of practical importance. Considering the sample mean of around 78\%, the results suggest a 4\% increase in the probability to stay in the starting line-up for the next match or a 14\% decrease in the probability of being replaced in the starting line-up for the next match. 

The findings are indicative of spillover effects in performance evaluations. A remaining concern associated with this interpretation involves the possibility that coworker quality, which likely influences coworker shooting performance, may exert an immediate impact on individual productivity, even after accounting for assortative matching, This concern is supported by \citeA{arcidiacono2017}, who present evidence of substantial spillover effects on individual productivity within teamwork settings. Despite our efforts to control for a comprehensive set of individual quality measures, team quality measures and other match-related variables to address assortative matching, it remains plausible that day-to-day fluctuations in coworker quality may jointly influence both coworker shooting performance and individual performance. Such a scenario would not allow us to rule out improvements in individual productivity as an alternative explanation for the observed results.\footnote{An alternative scenario might be so-called momentum effects, where coworker shooting performance does have an impact on individual productivity in the reminder of the match. While this does not invalidate the chosen identification approach, it would provide an alternative interpretation for the observed result.}

\begin{table}[H]
    \centering
    \begin{threeparttable}
        \caption{Robustness to alternative explanation}
        \begin{tabular}{lccc}
            \toprule
            & \multicolumn{3}{c}{Starting line-up next match}\\
            \cline{2-4} 
            \addlinespace
            & (1) & (2) & (3) \\  
            \midrule
            Coworker performance & 0.032$^{***}$ & 0.035$^{***}$ & 0.036$^{***}$\\   
            & (0.010) & (0.009) & (0.010)  \\ 
            \midrule
            Individual controls & $ \surd $ & $ \surd $ & $ \surd $ \\ 
            Team controls & $ \surd $ & $ \surd $ & $ \surd $ \\ 
            Match controls & $ \surd $ & $ \surd $ & $ \surd $ \\ 
            Coworker controls & $ \surd $ & & $ \surd $ \\
            Individual performance metrics & & $ \surd $ & $ \surd $  \\
            Observations & 58,968 & 58,968 & 58,968 \\  
            \midrule
        \end{tabular}
        \begin{tablenotes}
            \footnotesize
            \item \textbf{Notes:} *, **, and *** represents statistical significance at the 10\%, 5\%, and 1\% level respectively. Standard errors are clustered at the match level and are presented in parentheses. All effects are population averages (ATE). 
        \end{tablenotes}
        \label{table_estimation_results_main2}
    \end{threeparttable}
\end{table}

We consider the likelihood that fluctuations in coworker quality, which impact shooting performance, directly impacting the productivity of other players to be low in our setting. Shooting performance is mostly linked to a player's shooting skills. This type of skill is unlikely to have large spillover effects on productivity, and players that exhibit good shooting skills are not necessarily players with skills in other categories that contribute to enhancing the overall play of teammates. Despite this reasoning, we empirically address the potential issue that improvements in individual productivity explain the result in several ways and report the results in Table \ref{table_estimation_results_main2}.

In Column one, we add a set of control variables for coworker quality to the model. Specifically, coworker quality is controlled for similarly to the approach for player quality. This involves explicitly accounting for the skill levels of coworkers in the starting line-up across six game-relevant dimensions, including passing and dribbling, which are likely to generate the highest spillover effects on productivity. The result is insensitive to this model extension. 
To further exclude the possibility that individual productivity changes explain the observed result, we run an additional regression and include an extensive set of individual performance metrics in the match that should be able to adequately characterize individual productivity (Column two).\footnote{A list of performance metrics can be found in Table \ref{table_metrics1} and Table \ref{table_metrics2}.} Lastly, we combine both sets of control variables in our model (Column three). Importantly, the results remain virtually unaffected by the extensions to the estimation model. This result strengthens the interpretation of our findings as indications of spillover effects in performance evaluations. 

\begin{table}[h!]
    \centering
    \begin{threeparttable}
        \caption{Average effect of coworker performance on expert journalists' ratings}
        \begin{tabular}{lcccccc}
            \toprule
            & \multicolumn{6}{c}{Expert evaluation}\\
            \cline{2-7} 
            \addlinespace
            & (1) & (2) & (3) & (4) & (5) & (6) \\  
            \midrule
            Coworker performance & 0.432$^{***}$ & 0.391$^{***}$ & 0.333$^{***}$ & 0.298$^{***}$ & 0.339$^{***}$ & 0.337$^{***}$ \\   
            & (0.036) & (0.032) & (0.027)& (0.026) & (0.021) & (0.021) \\ \\
            Mean expert evaluation & 3.33 & 3.33 & 3.33 & 3.33 & 3.33 & 3.33\\  
            \midrule
            Individual controls & $ \surd $ & $ \surd $ & $ \surd $ & $ \surd $ & $ \surd $ & $ \surd $\\ 
            Team controls &  & $ \surd $ & $ \surd $ & $ \surd $ & $ \surd $ & $ \surd $\\ 
            Match controls &  &  & $ \surd $ & $ \surd $ & $ \surd $ & $ \surd $\\ 
            Coworker controls &  & & & $ \surd $ & & $ \surd $\\
            Individual performance metrics & &  & & & $ \surd $ & $ \surd $\\
            Observations & 58,968 & 58,968 & 58,968 & 58,968 & 58,968 & 58,968 \\  
            \midrule
        \end{tabular}
        \begin{tablenotes}
            \footnotesize
            \item \textbf{Notes:} *, **, and *** represents statistical significance at the 10 \%, 5 \%, and 1 \% level respectively. Standard errors are clustered at the match level and are presented in parentheses. All effects are population averages (ATE). For a full list of covariates included in each specification, see Table \ref{table_desc_stat2}. 
        \end{tablenotes}
            \label{table_estimation_results_exp}
    \end{threeparttable}

\end{table}

So far, we relied on manager decisions whether to field a player in the next match as our measure of performance evaluations. We now show results using an alternative measure of performance evaluation. Table \ref{table_estimation_results_exp} shows how coworker performance affects journalists' evaluations. While an obvious advantage of manager decisions is that they have direct career implications for players, expert journalists' ratings also have advantages.\footnote{Getting the chance to play more often and to show his skills is important for career advancement \cite{hoey2022}.} They are conducted by experts that are not part of the club. Therefore, they are free from strategic considerations about team success in the upcoming match as well as the availability of replacing players and more likely to reflect a "public perception" of individual performance. In line-with the results for manager decisions, we observe a substantial improvement in the evaluation a player receives from the expert journalist, equivalent to more than one-third of a full grade. The coefficient size corresponds to an approximate 10\% increase relative to the sample average grade.

To strengthen the validity of our results, we conduct two additional robustness checks. First, we examine the robustness of the findings to treatment outliers. In this robustness check, we exclude all observations where coworkers scored more than two goals. The results, shown in Table \ref{table_estimation_results_margin}, confirm the main result.\footnote{Additionally, we show all treatment comparisons (1 goal vs. 0 goals, 2 goals vs. 1 goal). For expert evaluations, positive and significant coefficients are observed for both comparisons. However, the impact of scoring two goals instead of one is larger than the effect of scoring one goal instead of zero goals across all specifications. Turning to manager decisions, a similar pattern is identified. Nevertheless, the effect of coworkers scoring one goal compared to those scoring zero goals is not significant in any specification.} Second, we demonstrate that the results are robust to the chosen estimator. We utilize a linear regression framework. The results, presented in Table \ref{table_femodel}, uphold our main findings. While the linear regression framework limits the ability to flexibly control for the extensive set of control variables, we also show results using individual or even individual-season fixed effects that address assortative matching based on observed and unobserved individual characteristics. Reassuringly, results using the set of control variables used in our main specification and results using individual or individual-season fixed effects are quantitatively very similar, strengthening the credibility of the condition independence assumption.

\subsection{Reference point dependence} \label{subsec: Nonlinear}

The previous section has shown that coworkers exhibit spillover effects on individual performance evaluations. We now turn towards understanding further circumstances that trigger spillover effects. This section investigates whether good or bad performances differentially initiate spillover effects. Specifically, we examine whether the coworker spillover effect is dependent on a reference point, demonstrating a differential impact below and above expectations.

Theoretical work has conceptualized expectations as reference points 
\cite{kHoszegi2006} and several empirical studies from the lab and the field provide evidence for reference point dependent behavior in various domains \cite{gneezy1997, pope2011, allen2017}. We test whether coworker performance below and above reference points exhibits a differential impact on a worker's performance evaluations. To do so, we expand our investigation by incorporating an additional method from causal machine learning. We adopt the non-parametric kernel method for continuous treatment proposed by \citeA{kennedy2017}. Instead of using a discrete treatment, the treatment variable is defined as the difference between the number of goals scored by coworkers and the number of expected goals coworkers were exposed to. In addition to understanding whether spillover effects are reference point dependent, this approach serves as an additional complementary estimation strategy to test the robustness of the main result.\footnote{In the first stage, a pseudo-outcome is constructed:

\begin{equation*}
	\xi(\pi,\mu) = \frac{Y-\mu(X,D)}{\pi(D|X)} \int \pi(D|x)dP(x) + \int \mu(x,D)dP(x),
\end{equation*}	

where $\mu(D|X)$ and $\mu(x, D)$, the nuisance functions, are estimated by a random forest
estimator \cite{breiman2001}. The pseudo-outcome $\xi(\pi, \mu)$ eliminates the influence of confounding factors and possesses the desirable characteristic of being doubly-robust. 
To estimate the average potential outcome for each treatment level, we employ non-parametric kernel regression on the pseudo-outcome. 
In determining the bandwidths of the kernels for this regression, we adopt a data-driven approach using a cross-validation method. }

\begin{figure}[h!]

      \caption{Spillover effects and reference point behavior}
      \label{fig: expected_goals}
    \begin{subfigure}{.49\textwidth}
      \centering
      \includegraphics[width=\textwidth]{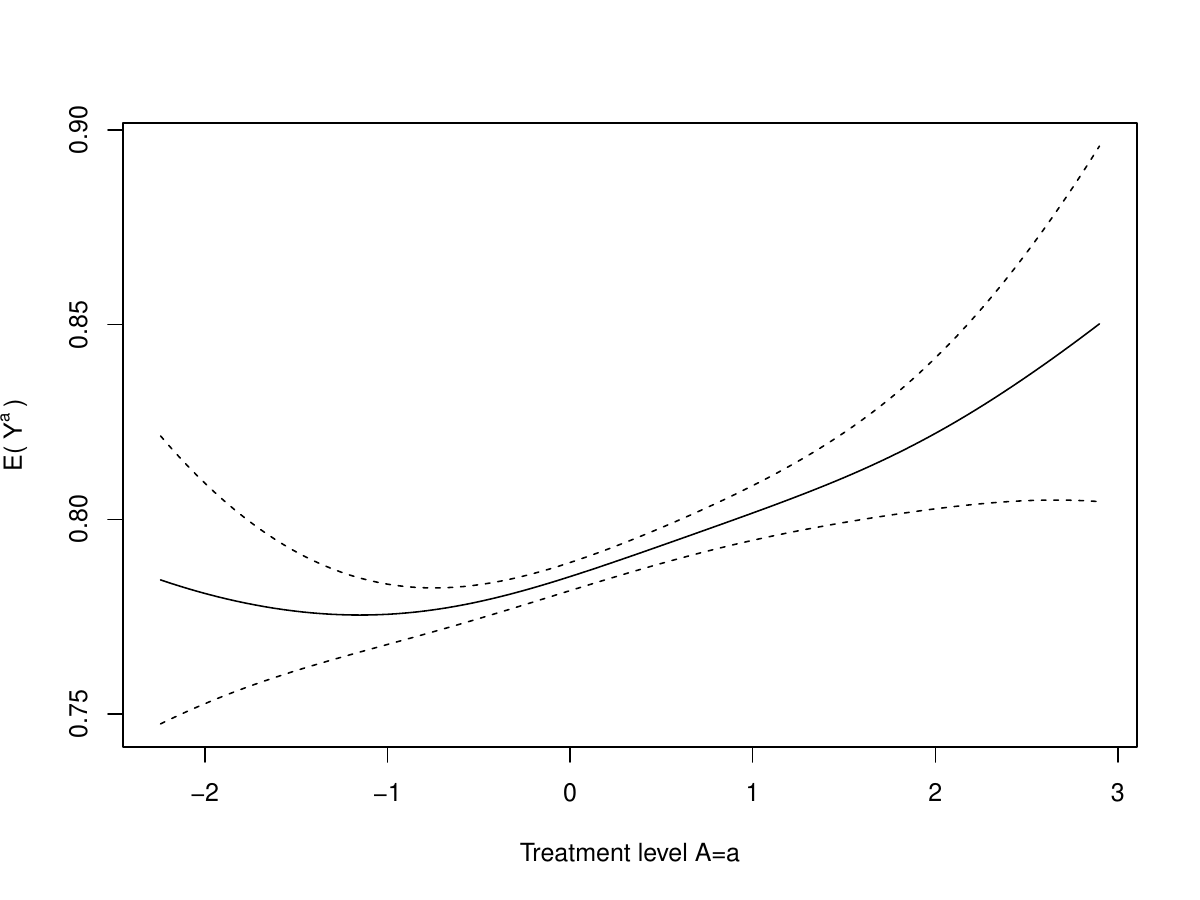}
      \caption{Manager}
      \label{fig: expected_goals_shots}
    \end{subfigure}
    \hfill
        \begin{subfigure}{.49\textwidth}
      \centering
      \includegraphics[width=\textwidth]{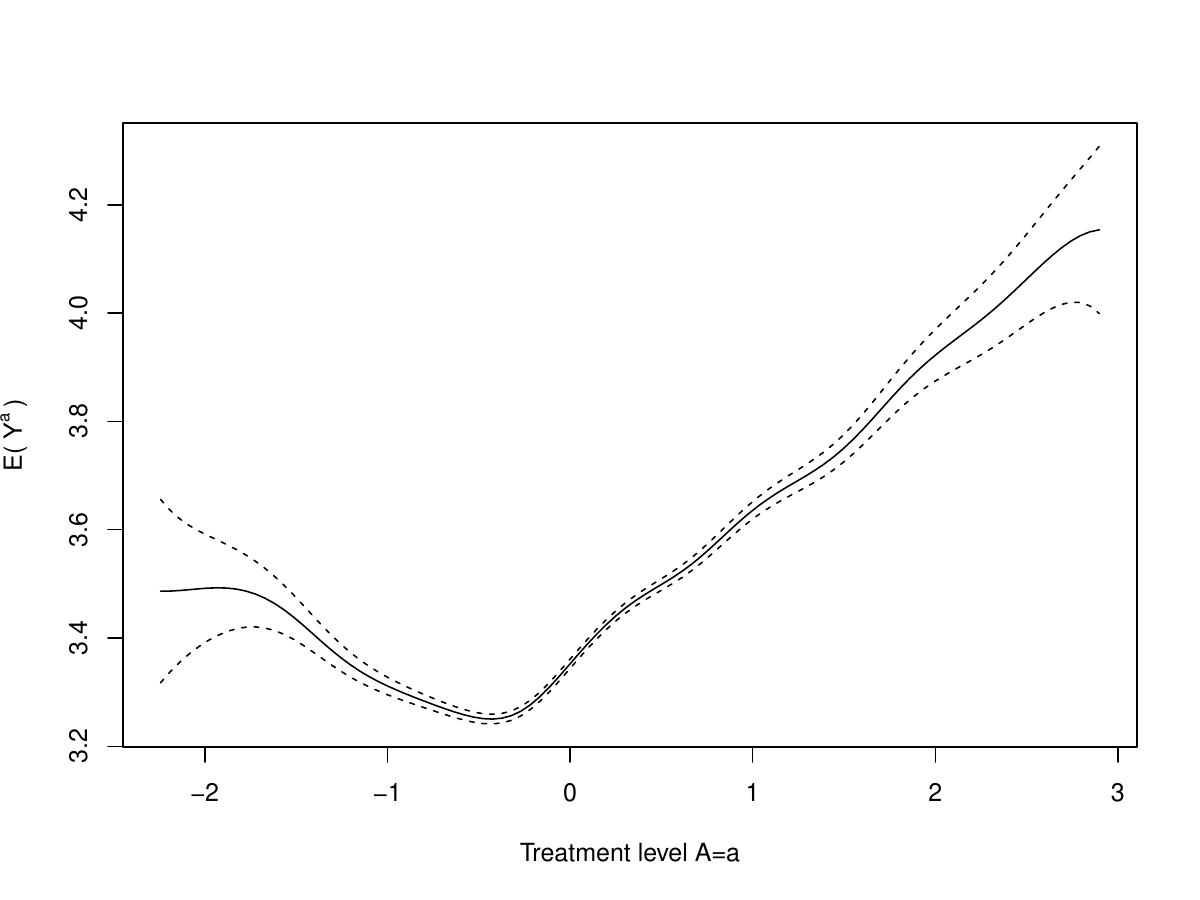}
      \caption{Expert}
      \label{fig: expected_goals_ind}
    \end{subfigure}    
	\footnotesize \textbf{Notes:} Non-parametric kernel regression based on Table \ref{table_estimation_results_main}, Column one. Subsample without treatment outliers (99\%). The solid line shows the performance evaluation for different levels of coworker performance. Dashed lines show confidence intervals around the smoothed parameter.\label{fig_conttreat_main}
\end{figure}

The estimates are presented in Figure \ref{fig_conttreat_main}. We plot the estimated average potential outcomes for each treatment level. For two different treatment levels $D = d_1$ and $D = d_0$ (on the x-axis), the treatment effect can be calculated as $\theta(d_1, d_0) = \frac{E(Y(D = d_1)) - E(Y(D = d_0))}{d_1 - d_0}$. The treatment intensity in this example is $d_1 - d_0$, with $d_0$ being the treatment level from which the intensity is evaluated.

We observe the following key findings: First, in both subfigures, we notice some non-linearity. Positive deviations from the expectation have a substantial positive impact on coworkers. Negative deviations are hardly harmful for coworkers. This holds true for both manager decisions and expert evaluations.
Second, manager decisions exhibit a rather monotone relationship across the entire distribution of the treatment variable, with a rather flat relationship below expectations and a positive relationship above expectations. However, this differs for expert evaluations. Workers receive the lowest performance evaluation if coworkers perform marginally below expectation. If they perform substantially below expectation, workers receive a smaller penalty. This result may stem from coworker performance showing very obvious mistakes that can be easily associated with them.

\subsection{Effect heterogeneity} \label{subsec: Heterogeneity}

\subsubsection{Heterogeneity in worker and firm characteristics} \label{subsubsec: Heterogeneity1}

In this subsection, we concentrate on identifying which group of workers is most and least affected by spillover effects. To start uncovering effect heterogeneity, we present Group Average Treatment Effects (GATEs) for four characteristics at the individual and firm level that have implications beyond the sports industry.

\begin{figure}[h!]

      \caption{GATEs for different player characteristics}
      \label{fig: expected_goals}
    \begin{subfigure}{.49\textwidth}
      \centering
      \includegraphics[width=\textwidth]{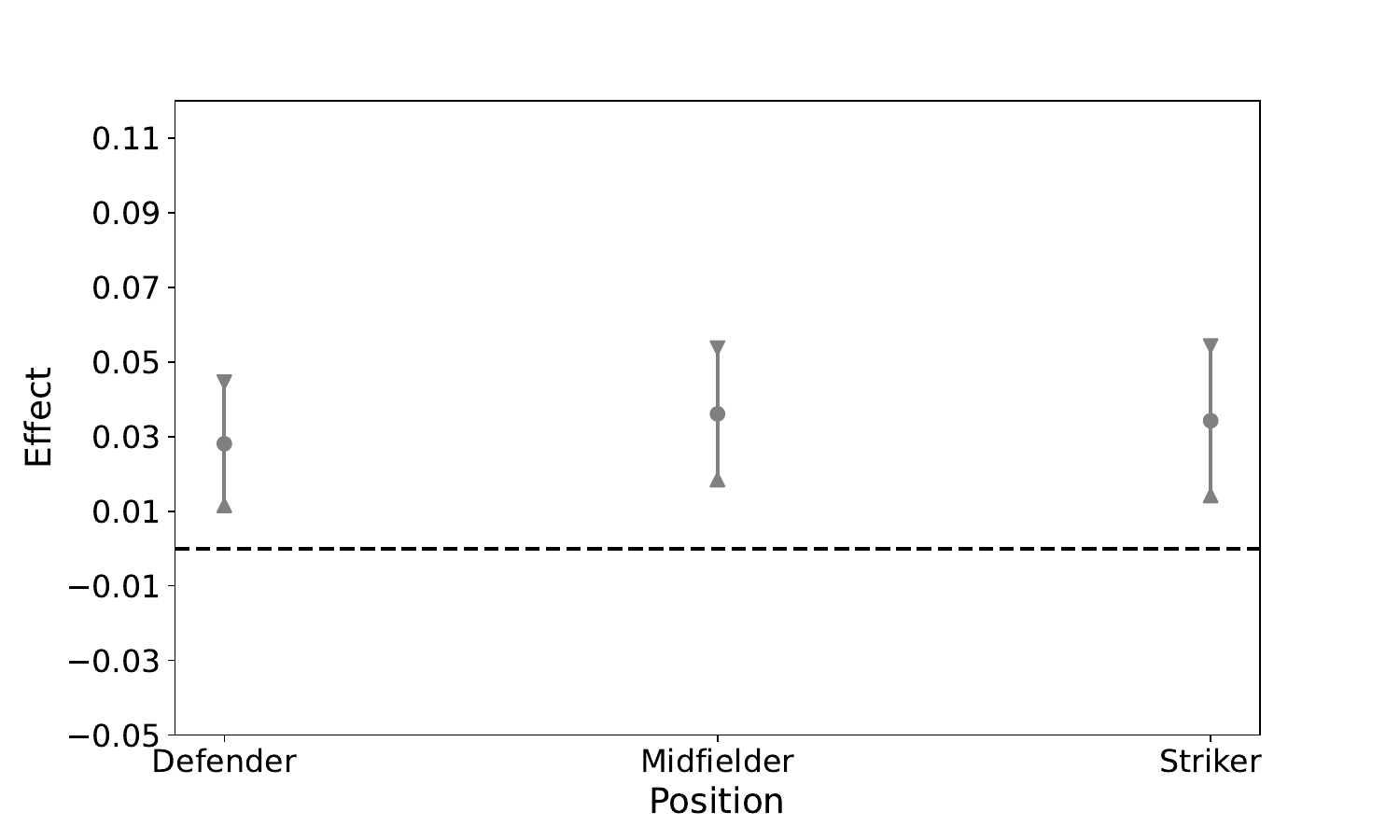}
      \caption{Manager}
      \label{fig: expected_goals_shots}
    \end{subfigure}
    \hfill
        \begin{subfigure}{.49\textwidth}
      \centering
      \includegraphics[width=\textwidth]{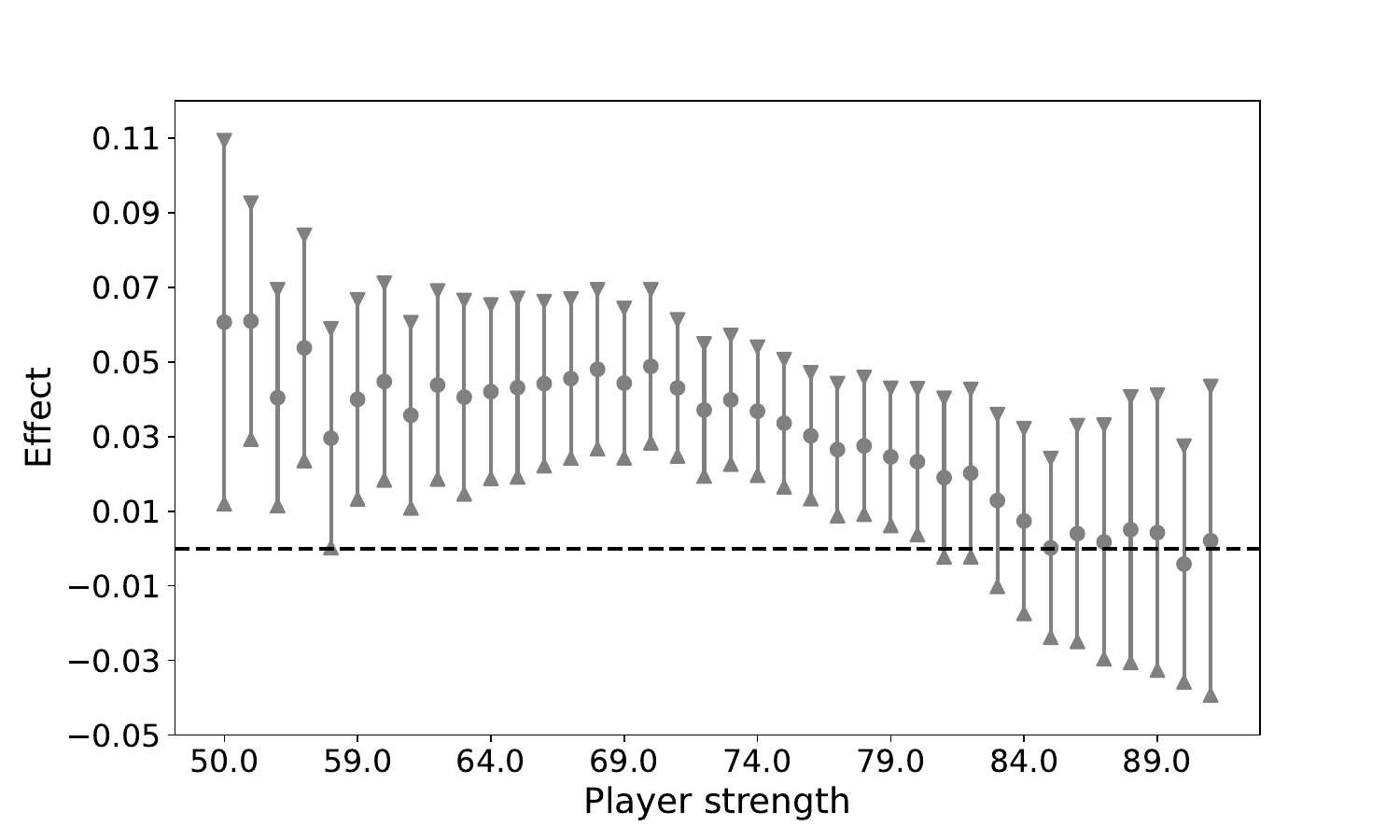}
      \caption{Manager}
      \label{fig: expected_goals_ind}
    \end{subfigure}    
    \begin{subfigure}{.49\textwidth}
      \centering
      \includegraphics[width=\textwidth]{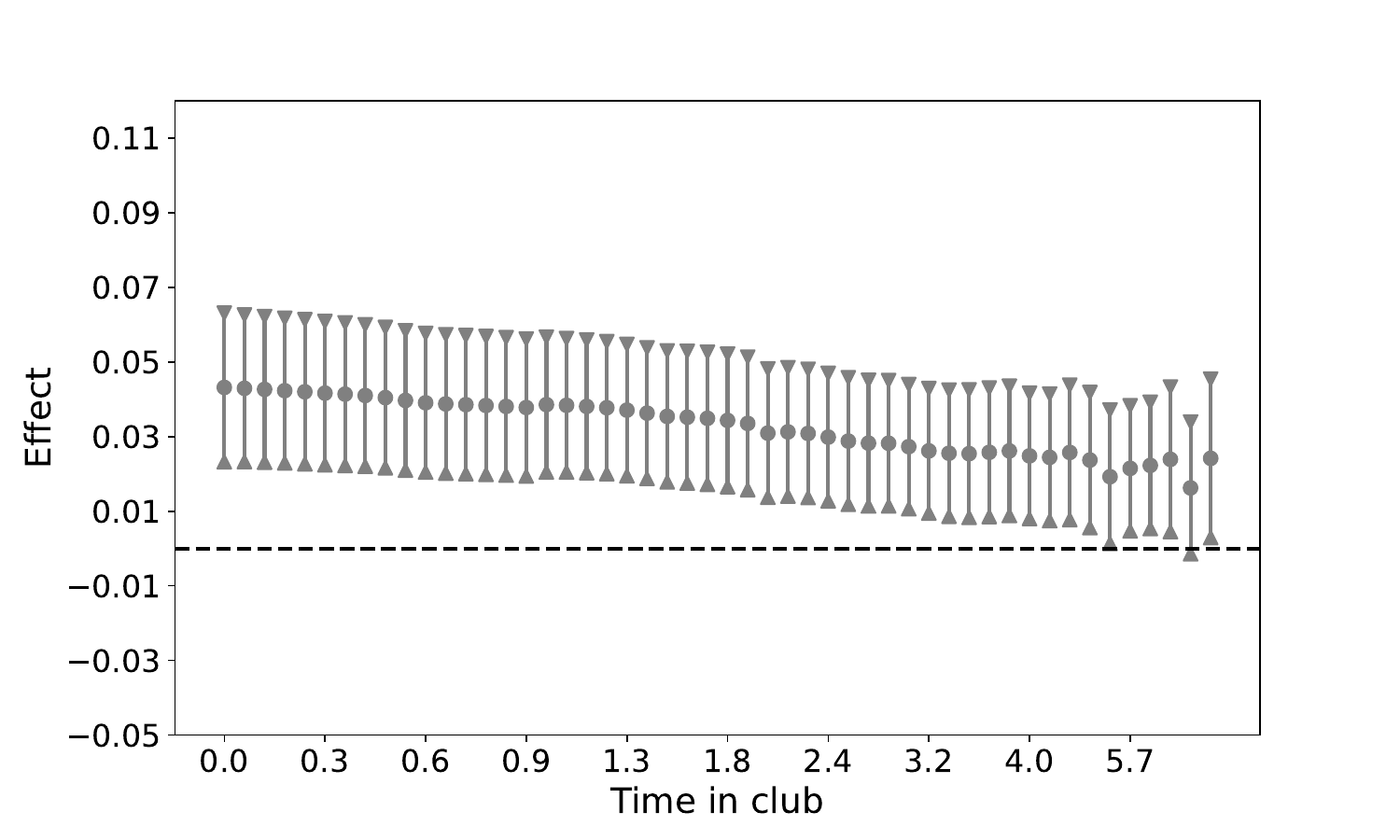}
      \caption{Manager}
      \label{fig: expected_goals_shots}
    \end{subfigure}
    \hfill
        \begin{subfigure}{.49\textwidth}
      \centering
      \includegraphics[width=\textwidth]{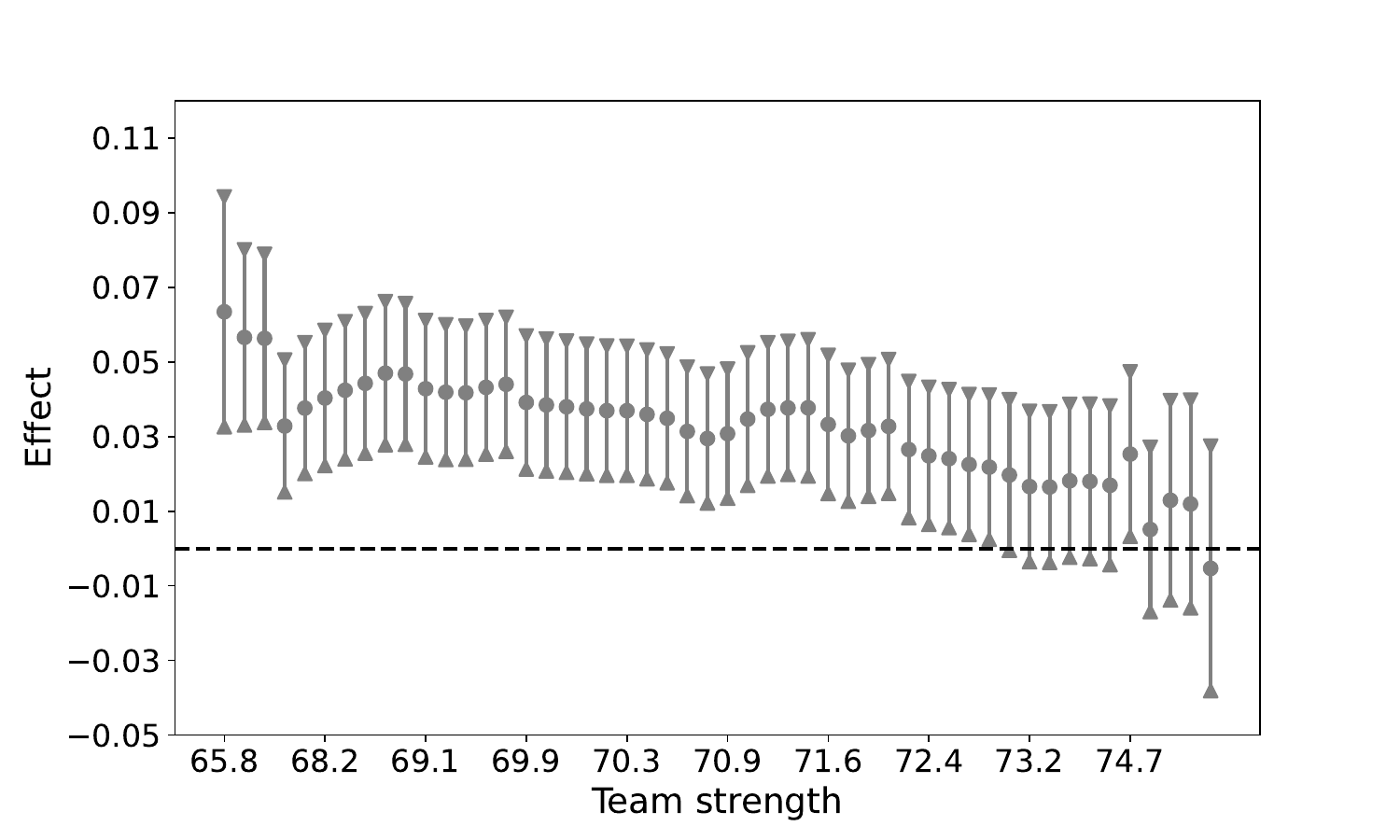}
      \caption{Manager}
      \label{fig: expected_goals_ind}
    \end{subfigure}    
	\footnotesize \textbf{Notes:} The vertical axis denotes the respective GATE and its 90\% confidence interval for the comparison of coworkers scoring two goals, relatively to coworkers scoring zero goals. The measure for player strength and team strength is the video game measure (0 to 100).  \label{fig_cates}
\end{figure}

First, in most organizations and teamwork settings, workers are assigned different sub-tasks aligned with their primary responsibilities, known to the manager. Shooting performance, for instance, is particularly crucial for strikers, while defenders primarily focus on preventing the opposing team from scoring. We investigate whether task responsibility affects the spillover effect in performance evaluations. Panel (a) of Figure \ref{fig_cates} shows how the effect of coworker performance varies based on the player's position, relying on pre-season player classifications of their main positions collected by Opta. We find that the effect of coworker performance on the manager's fielding decision is significantly different from zero for players in all positions. The estimated GATEs are neither statistically different from the average treatment effect nor from each other.\footnote{While we also have access to the precise position on matchday, we abstain from using it due to the challenges of classifying players based on fluid positions and tactical changes during a match. In results not shown in the paper, we also conducted our main analysis including goalkeepers. There is no effect on the probability of staying in the starting line-up in the next match, which is plausible given the generally limited variation in goalkeeper positions.}

Second, we explore whether more or less skilled players are differentially affected. Player strength significantly influences the manager's line-up decisions, suggesting that more skilled players should be less susceptible to variation in coworker performance compared to weaker players, especially concerning manager decisions. As depicted in Panel (b) in Figure \ref{fig_cates}, we observe a decreasing effect size with player strength. However, it's noteworthy that the vast majority of GATEs are positive, with only star players being seemingly unaffected.

Third, we investigate heterogeneity concerning the time a player has been with the club. The longer a player has been with the club, and the more familiar the manager is with the player, the greater the trust and knowledge about the player's quality. Panel (c) shows that almost all GATEs are positive and statistically different from zero. We also observe effect sizes ranging from 4 percentage points for new players to around 2 percentage points for players who have been with the club for a substantial period. 


Lastly, we investigate whether the experience of the manager or the quality of the club matter for the importance of a coworker bias. Panel (d) therefore shows GATEs with respect to the average quality of the whole team. Spillover effects are substantially larger for lower levels of team quality, thus for teams that are less frequently used to team success.

We also estimate effect heterogeneity for the alternative outcome variable. For brevity reasons, results from heterogeneity analysis for journalists' ratings are subsumed in Appendix D. In line with the idea that journalists' views reflect a more public perception of players and the role of players in a team matter less for journalists, we find no heterogeneity in the effect size with regard to player strength and tenure level. However, we observe some heterogeneity by the position of players. The effect of coworker performance is significantly different from zero for players in each position, albeit the effect of coworker shooting performance is notably smaller for defenders compared to strikers.

\subsubsection{Fine-grained heterogeneity} \label{subsubsec: Heterogeneity2}

In the preceding section, we explored the extent to which effect heterogeneity is linked to a set of relevant firm and worker characteristics. To complement the heterogeneity analysis, we use data driven approaches to detect effect heterogeneity and to characterize groups of players that are most and least affected by spillover effects.\footnote{Panel A in Figure \ref{fig_iate} provides a visual representation of the distribution of Individualized Average Treatment Effects (IATEs) when comparing coworkers scoring two goals with those scoring zero goals. Additionally, Panel B in Figure \ref{fig_iate} presents the sorted effects along with a 90\%-confidence interval based on the estimated standard errors. Two main observations arise from these figures. First, the majority of the IATEs are positive for both outcome variables. Second, there is significant variation in the effects, indicating meaningful heterogeneity in the treatment effects. Complementary results for journalists' ratings, see Figure \ref{fig_iate2}.} Detecting additional patterns of heterogeneity is challenging without some additional structure. Therefore, we show the findings from an optimal policy simulation. While our goal is not to learn treatment assignment policies, since the treatment cannot be modified, an optimal policy simulation approach offers the advantage of producing yet an interpretable characterization of the groups that benefit most and least from the treatment with regard to a joint characterization based on a set of predefined features.

\begin{table}[h!]
    \centering
    \begin{threeparttable}
        \caption{Most and least affected group of players - policy simulation}
        \begin{tabular}{l *{3}{p{1.8cm}}}
            \toprule
            & \multicolumn{3}{c}{Treatment shares in \%}\\  
            \midrule
             & 0 Goals & 1 Goal & 2 Goals \\
             \midrule
            Observed & 33.69     & 34.13    & 32.18 \\
            Random  & 32.65     & 35.14     & 32.21 \\
            Decision tree 3 levels & 5.01     & 0     & 94.99 \\
            \midrule
             & \multicolumn{3}{c}{Mean of features in assignment of decision tree}\\
             \midrule
            Player quality & 85.83   & –     & 74.51 \\ 
            Time in club & 4.35   & –      & 2.33 \\ 
            German & 0.44   & –      & 0.43 \\ 
            Manager experience & 469.33   & –      & 136.54 \\ 
            Team strength & 76.72   & –      & 70.91 \\ 
            Position & 2.86   & –      & 2.75 \\ 
            \midrule
        \end{tabular}
        \begin{tablenotes}
            \footnotesize
            \item \textbf{Notes:} This table provides treatment shares in percentage for different outcome categories in a policy simulation related to "Starting line-up." The second part of the table shows the mean of features for each category.
        \end{tablenotes}
        \label{table_policytree_Startelf1}
    \end{threeparttable}
\end{table}


We derive policy rules based on a modification of the approach by \citeA{zhou2023}, implemented in decision trees of depth 3 (8 strata) in the \textit{mcf} package.
Table \ref{table_policytree_Startelf1} displays the description of the simulated allocation rule and the shares of the population allocated to the different treatments (in \%). We also depict the observed shares and the shares that would have been observed under random assignment. A decision tree of depth 3 assigns 95\% of observations to the highest treatment dose, while 5 \% are assigned to the zero treatment. This is reassuring given that the vast majority of IATEs and GATEs are above zero.\footnote{The fact that the assigned shares under random assignment and the observed shares are almost identical reveals that managers do not seem to base their fielding decisions on the effect heterogeneity of spillover effects. This is reassuring, given that managers do not intend to keep the same starting line-up, but seek to maximize success.}

In the lower part of Table \ref{table_policytree_Startelf1}, we present the average characteristics of features used in constructing the decision tree, categorized by assigned treatment status. Consistent with the previous sections, the decision tree tends to assign high-quality players from stronger teams, with more club tenure, and under the guidance of experienced managers to the no-goal treatment. This aligns closely with our earlier findings. Table \ref{table_policytree_Startelf} shows that only two out of eight final leaves are assigned to the no-goal treatment, with the remaining leaves assigned to the two-goal treatment. The assignment rules mirror the patterns observed in Section \ref{subsubsec: Heterogeneity1}. Notably, some degree of substitutability between manager experience and player quality emerges, with the first leaf featuring highly experienced managers and high player quality, while the second leaf showcases very high player quality, albeit with lower manager experience.\footnote{Complementing results using a k-means++ clustering approach can be found in Table \ref{table_cluster_Startelf}. The results align with our previous results. Spillover effects range from 0.12 in the highest cluster to 0 in the lowest cluster.} 

For our alternative outcome variable, the optimal policy simulation does not result in a useful characterization of players that are most and least affected by spillover effects, since all observations are assigned to the two goal treatment. Therefore, we depict the dependence of the effects on covariates using k-means++ clustering \cite{arthur2007k}. The results are shown in Table \ref{table_cluster_Expert} in Appendix \ref{subsec: Appendix D} along with a detailed description. 

The analysis reveals substantial heterogeneity in the degree of spillover effects. The spillover effects on journalists' evaluations range from 0.18 in the least affected cluster to 0.51 in the most affected cluster. To facilitate interpretation, consider one example. A 0.51 increase in the evaluation received by journalists is roughly equivalent to 0.5 times the effect of scoring one goal oneself, given the faced chances. The size of the spillover effect is, thus, one-quarter of the effect of one's own performance in the same task.\footnote{Results used for this comparison are shown in Appendix C Table \ref{table_estimation_results_own}.} The characterization of clusters aligns with our results from the previous section. Manager and team strength matter less compared to the manager's decision. The group benefiting the most from spillover effects is characterized by players with high offensive skills and a good team performance (xG), while player exhibiting high defensive skills are prevalent in the group of player who benefit least from their coworkers.

\subsection{Proximity to coworkers} \label{subsec: Tasksandcloseness}

In Section \ref{subsubsec: Heterogeneity1}, we explored whether players who are responsible for a similar task (goal scoring is the primary task of offensive players) are differentially affected than players who's primary task is to prevent goals (defensive players). While we identified a larger impact on offensive players in journalists' evaluations, no such variation was observed for manager decisions. This section extends the analysis. First, we ask whether players, who's main task is scoring goals, exhibit higher spillover effects on other workers than workers who's main task is different (defensive task). Then, we ask whether the proximity to coworkers (shared responsibilities), rather than the own or the coworker responsibilities, affect the size of the spillover effect.


Tables \ref{table_estimation_results_main3} and \ref{table_estimation_results_main4} present separate analyses for manager and expert evaluations. We calculate coworkers' shooting performance for different subgroups based on their main positions and estimate the effects on performance evaluations, considering the player's own position. We focus on strikers and defenders, since task responsibilities are more straightforward compared to midfielders. Due to common support issues, the sample is restricted to observations where coworkers score zero or one goal. 

Table \ref{table_estimation_results_main3} reveals no differential impact of coworker shooting performance depending on the main responsibilities of coworkers. Neither are shared responsibilities associated with a heterogeneous effects size. This may suggest that the primary driver for spillover effects is team performance.
For journalists' ratings (Table \ref{table_estimation_results_main4}), we find that shooting performance of strikers has a stronger impact compared to shooting performance of defenders (Column one). Thus, coworkers main responsibilities matter for the size of the spillover effect they exhibit on coworkers. We also find that the effect of coworkers shooting performance always tend to be stronger for strikers, independent of the position of the player who provided the performance. Proximity to coworkers, shared responsibilities, are thus not associated with an additional premium.

\section{Career effects} \label{sec: Career effects}

To explore the long-term consequences of coworker performance, we investigate whether aggregates of our measure of shooting performance are associated with player assessments in subsequent periods. To do so, we aggregate the main dataset at the player-by-season level, resulting in approximately 3,900 player-season observations. Our measure of coworker performance at the season level is the aggregated number of goals scored by coworkers and the aggregated xG coworkers were exposed to in a given season $t$. We then examine the relationship between these aggregated coworker performance variables and expert-based player evaluations in the subsequent season ($t+1$) and the season after next ($t+2$), using data from \textit{transfermarkt.de}. This source provides expert-based evaluations that are highly correlated with transfer fees, offering a valuable indicator of a player's perceived value in the football market \cite{prockl2018, coates2022}. An important advantage of these evaluations compared to transfer fees is that they are observed independent of a player changing the club, which avoids selection issues. Furthermore, in contrast to wage information, they are unlikely to depend on players bargaining power or other sources apart from the actual or perceived productivity.

\begin{table}[h!]
    \caption{Career perspective}
    \centering
    \begin{threeparttable}
        \begin{tabular}{lcccc}
            \toprule
            & \multicolumn{4}{c}{Player evaluation (market value)}\\  
             & \cline{1-4} 
            \addlinespace
            & \multicolumn{2}{c}{$t + 1$} & \multicolumn{2}{c}{$t + 2$}\\ 
            &   \cline{1-4} 
            \addlinespace
            Coworker performance         & 75,659$^{***}$ & 75,204$^{**}$ & 127,784$^{***}$  & 91,666$^{***}$\\   
            & (23,795)   & (30,738)     & (37,116) & (34,039) \\  
            \midrule
            Individual controls   & $ \surd $       & $ \surd $     & $ \surd $    & $ \surd $ \\ 
            Team controls    & $ \surd $   & $ \surd $  &  $ \surd $       & $ \surd $  \\ 
            Season Fixed effects   &       & $ \surd $    &  & $ \surd $ \\ 
            Individual Fixed effects   &   & $ \surd $   &     & $ \surd $  \\ 
            Observations   & 3903          & 3903 & 3903  & 3903\\  
            \midrule
        \end{tabular}
        \begin{tablenotes}
            \footnotesize
            \item \textbf{Notes:}  Linear regression. Standard errors are clustered at the team-season level. $t$ is the current season, $t+1$ refers to the next season and $t+2$ refers to the season after the next season. Individual controls are the same as used in \ref{table_estimation_results_main}. Team controls are proxies for team quality. *, **, and *** represents statistical significance at the 10\%, 5\%, and 1\%.
        \end{tablenotes}
    \end{threeparttable}
    \label{table_estimation_chap6}
\end{table}

Relating coworker shooting performance over a season to player assessments in subsequent years is complicated by endogeneity issues. To get as close as possible towards a causal interpretation of the results, we address the issue of selection into particular clubs by incorporating various control variables and fixed effects. Table \ref{table_estimation_chap6} presents the results for two specifications. In the first specification, we control for the same set of individual and team characteristics as presented in Table \ref{table_estimation_results_main}. The second specification adds season fixed-effects and, more crucially, individual fixed-effects. These individual fixed-effects help to account for any differences between players at the player level. In all specifications, we also control for the number of xG coworkers are exposed to in a season.

The findings indicate a significant positive association between coworker performance and the market value evaluations of workers in subsequent seasons. Specifically, for each goal scored by coworkers, given the chance quality they were exposed to, a worker's market value increases by 75,000 Euros in the subsequent season and by 90,000 Euros in the season after the next. Given that the average market value of all players in our sample in 2010 is approximately 4 million Euros, ten more goals scored by coworkers (equivalent to less than one in every third match) result in an almost 20\% increase in the market value of a worker in the next period relative to the sample mean.

\section{Conclusion}  \label{sec: Conclusion}

Coworkers exert considerable influence on their peers, impacting productivity and effort provision \cite{arcidiacono2017, cohen2023effort}. Moreover, exposure to high-quality peers contributes positively to earnings and subsequent career outcomes \cite{jarosch2021}. While existing explanations focus on learning from peers leading to skill acquisition and enhanced work ethics \cite{amodio2023b, nix2020}, this paper introduces an additional and potentially widespread channel through which coworkers shape the career trajectories of their peers.

Modern organizations extensively utilize teamwork to harness the advantages of specialization \cite{deming2017}. However, teamwork introduces challenges, particularly in isolating individual contributions within the context of team performance. This complexity creates the potential for spillover effects in performance evaluations, where coworker performance causally influences individual performance assessments, extending beyond coworkers' impact on individual productivity. Considering the pivotal role of performance evaluations in shaping careers, understanding the conditions and extent to which these evaluations depend on coworker performance becomes crucial.

We examine spillover effects in performance evaluations using highly detailed individual data from the professional soccer industry. Leveraging player tracking data, we identify situations where coworker performance appears as good as random, while controlling for a comprehensive and relevant set of variables to account for assortative matching of workers. Methods from causal machine learning allow estimating average and heterogeneous treatment effects. 

Our results indicate significant spillover effects in performance evaluations. Managers' decisions to field players and expert evaluations are notably influenced by coworker performance, with substantial effects. The effect size is equivalent to one-third of the effect of the own performance in the same task and reduces the likelihood to be replaced in the starting line-up by up to 13\% relative to the sample mean. Also, third party evaluation of performance, are impacted, showing an 11\% change relative to the control mean. We further characterize groups of players and circumstances that trigger spillover effects. Interestingly, spillover effects exhibit a reference point dependent behavior. Positive deviations from a reference point create large positive spillover effects, while negative deviations do not harm coworker evaluations. Furthermore, we find that less skilled worker, or worker with limited tenure in the club, are most effected by spillover effects in manager decisions. However, the vast majority of players are affected. For third-party experts, we document that shared responsibilities are driving effect heterogeneity. 

Given the pronounced spillover effects observed in our setting, we believe such effects are likely to be even more prevalent in other organizational contexts. The visibility of soccer performance, with players tracked by cameras and detailed statistics available, makes it an environment where coworker effects are noticeable. We posit that these spillover effects are widespread in situations where actions are less monitored, and decision-makers may not have strong incentives to make optimal choices. This has significant implications, introducing inefficiencies and inequities in reward allocation and promotions. Moreover, it has substantial consequences for the career trajectories of workers.

\clearpage
\onehalfspacing
\bibliography{a_lit}
\bibliographystyle{ecta}
\doublespacing
\clearpage

\newpage
\begin{appendices}

\doublespacing

\counterwithin{table}{section}
\counterwithin{figure}{section}

\section{Data and Descriptive Statistics} \label{subsec: Appendix A}

In this section we present details about our data and descriptive statistics.

\begin{figure}[h!]

      \caption{Examples of Fifa player evaluation cards}
      \label{fig: fifa_eval}
    \begin{subfigure}{.3\textwidth}
      \centering
      \includegraphics[width=\textwidth]{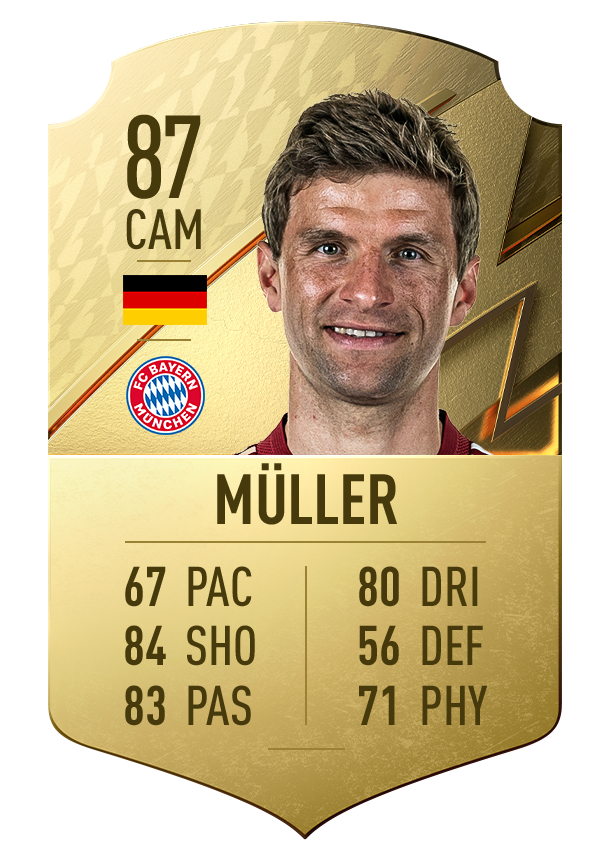}
      \caption{Thomas Müller}
    \end{subfigure}
    \hfill
        \begin{subfigure}{.3\textwidth}
      \centering
      \includegraphics[width=\textwidth]{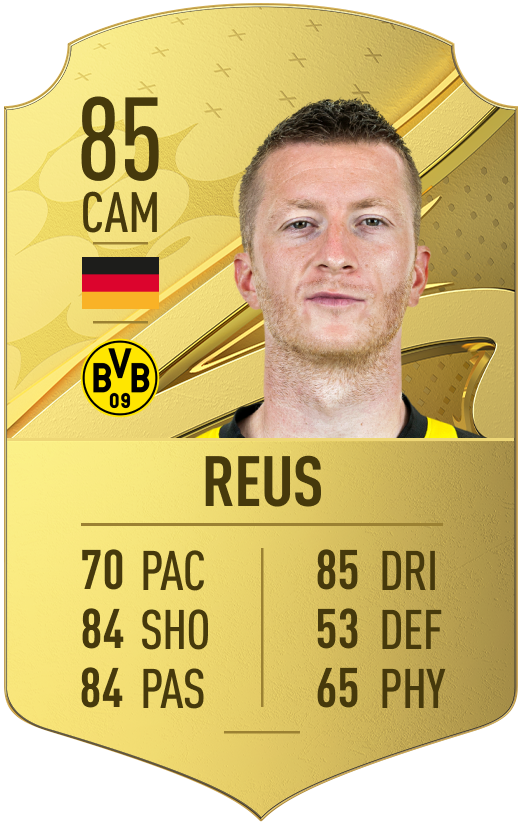}
      \caption{Marco Reus}
    \end{subfigure}     
    \hfill
        \begin{subfigure}{.3\textwidth}
      \centering
      \includegraphics[width=\textwidth]{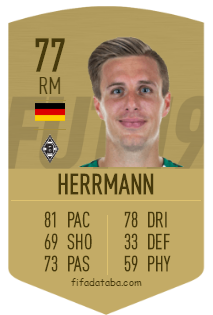}
      \caption{Patrick Herrmann}
    \end{subfigure}   
	\\ \footnotesize \textbf{Notes:} PAC = Pace; SHO = Shooting; PAS = Passing; DRI = Dribbling; DEF = Defensive skills; PHY = Physical skills.
\end{figure}

\clearpage

\begin{figure}[h!]
    \caption{Example of Newspaper evaluation}
    \begin{center}
	\includegraphics[width=0.5\linewidth]{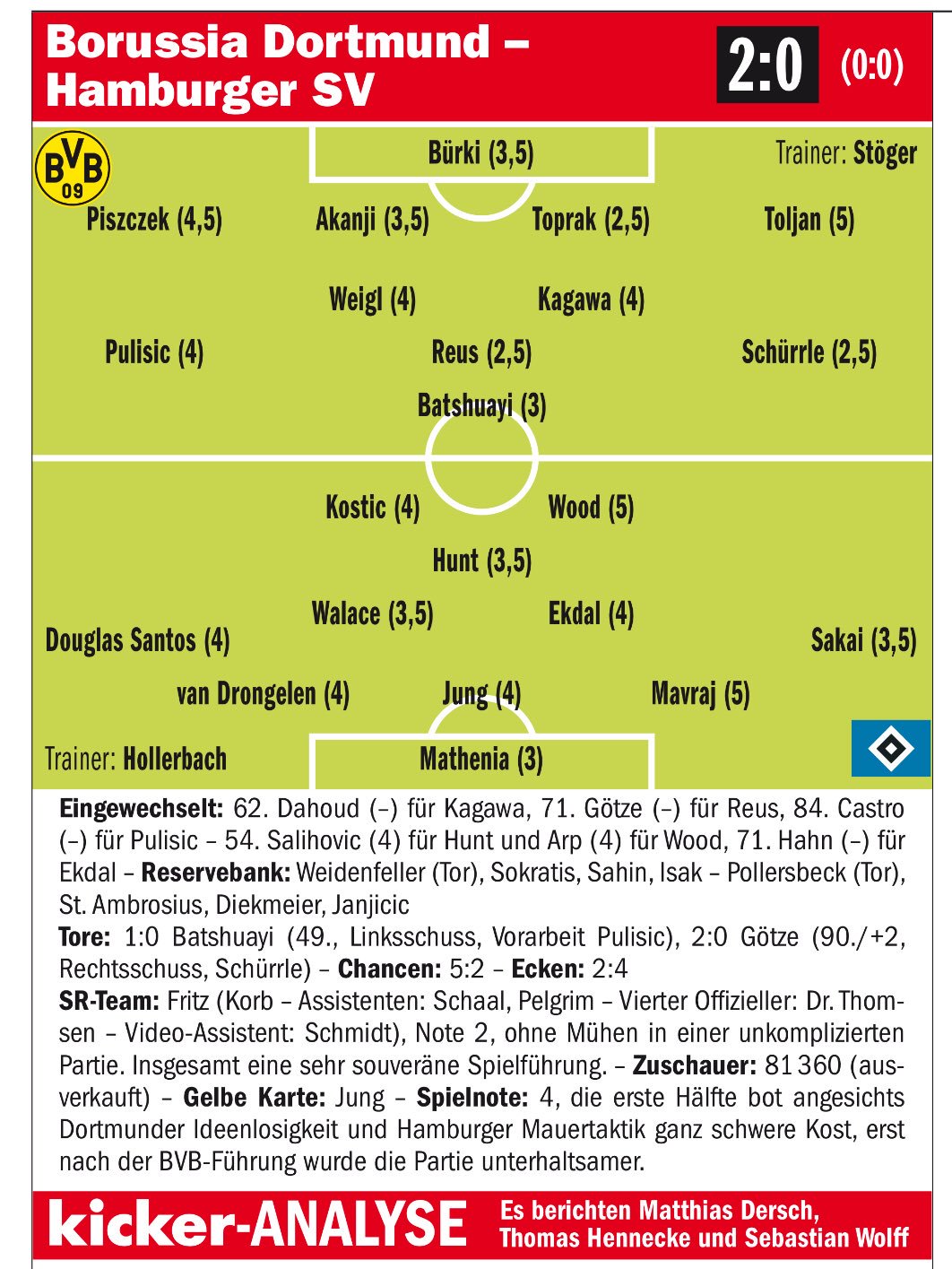}         
    \end{center}
	\footnotesize \textbf{Notes:} Player evaluations are a topic of public debate (https://www.sueddeutsche.de/sport/fussball-bundesliga-noten-vielleicht-reicht-auch-mal-eine-fuenf-1.141370). For more details on the construction of evaluations, see https://www.news.de/sport/855093240/die-zeugnisse-der-profis/1/.\label{fig: kicker}
\end{figure}

\begin{table}[H]
    \centering
    \begin{threeparttable}
    \caption{Descriptive Statistics by Treatment Status} 
\scriptsize
\begin{tabular}{lcccccc}
	\tabularnewline 
    \toprule
	& \multicolumn{6}{c}{Goals scored by coworkers}  \\  
     \midrule 
	& \multicolumn{2}{c}{0}             & \multicolumn{2}{c}{1}              & \multicolumn{2}{c}{2}   \\  
	\midrule
	\emph{Outcome Variables}\\ \\
	Starting line-up next match &   0.766 & (0.423) &   0.780 & (0.414) &   0.813 & (0.390) \\
 Journalist rating &   3.083 & (0.927) &   3.374 & (0.907) &   3.800 & (0.889) \\   
    \midrule
	
	\emph{Covariates}\\ \\

    \emph{Chance quality}\\
expected goals coworker &       0.704 &       (0.474) &       1.087 &       (0.573) &       1.733 &        (0.792) \\ \\
\emph{Individual controls}\\
        Position &   2.819 &  (0.742) &   2.766 &  (0.722) &   2.699 &  (0.693) \\
             Foot &   1.362 &  (0.668) &   1.351 &  (0.657) &   1.344 &  (0.631) \\
     Time in club &   2.309 &  (2.162) &   2.407 &  (2.234) &   2.603 &  (2.413) \\
              Age &  26.170 &  (3.683) &  26.140 &  (3.691) &  26.057 &  (3.736) \\
           German &   0.440 &  (0.496) &   0.439 &  (0.496) &   0.433 &  (0.496) \\
            Market value (ln) &  15.058 &  (1.080) &  15.173 &  (1.107) &  15.404 &  (1.187) \\
          Player strength (Fifa) &  74.334 &  (4.803) &  74.835 &  (4.914) &  75.760 &  (5.363) \\
        Defensive Skills (Fifa) &  57.895 & (18.968) &  59.455 & (18.793) &  61.725 & (18.470) \\
        Pace Skills (Fifa) &  71.340 & (10.379) &  71.522 & (10.281) &  71.960 & (10.454) \\
        Passing Skills (Fifa) &  65.780 &  (8.729) &  66.430 &  (8.782) &  67.535 &  (9.078) \\
        Physical Skills (Fifa) &  70.857 &  (7.535) &  71.061 &  (7.498) &  71.328 &  (7.635) \\
        Shooting Skills (Fifa) &  59.735 & (14.364) &  59.419 & (14.363) &  59.312 & (14.339) \\
        Dribling Skills (Fifa) &  69.197 &  (9.960) &  69.557 & (10.033) &  70.171 & (10.450) \\ \\
    \emph{Team controls}\\
    Days since previous match &   7.788 &   (6.059) &   7.489 &   (5.572) &   7.491 &   (5.824) \\
       Days to next match &   7.499 &   (5.090) &   7.696 &   (5.933) &   7.173 &   (4.948) \\
         Upcoming Cup match &   0.125 &   (0.331) &   0.155 &   (0.362) &   0.198 &   (0.398) \\
          Following Cup match &   0.145 &   (0.352) &   0.153 &   (0.360) &   0.202 &   (0.402) \\
         Relegated last season &   0.152 &   (0.359) &   0.125 &   (0.331) &   0.084 &   (0.277) \\
   Rank last season &  10.353 &   (4.985) &   9.640 &   (5.091) &   8.385 &   (5.310) \\
       Number of previous clubs (manager) &   2.379 &   (1.984) &   2.413 &   (2.006) &   2.608 &   (2.142) \\
    Number of League Matches (manager) & 119.149 & (136.819) & 121.681 & (138.139) & 135.330 & (148.318) \\
      Number of Top-5 matches (manager) & 137.803 & (158.349) & 145.770 & (168.444) & 174.753 & (196.653) \\
     Average market value squad (ln) &  14.869 &   (0.713) &  14.994 &   (0.744) &  15.239 &   (0.838) \\
    Std. market value squad (ln) &  14.798 &   (0.834) &  14.941 &   (0.850) &  15.209 &   (0.929) \\
        Average market value Top 11 (ln) &  14.539 &   (0.760) &  14.670 &   (0.791) &  14.930 &   (0.881) \\
       Std. market value Top 11 (ln) &  14.607 &   (0.911) &  14.747 &   (0.912) &  15.010 &   (0.965) \\
   Average squad strength (Fifa) &  70.618 &   (2.375) &  71.058 &   (2.481) &  71.763 &   (2.766) \\
  Std. squad strength (Fifa) &   6.492 &   (1.350) &   6.667 &   (1.434) &   7.067 &   (1.645) \\   \\
    \emph{Match controls}\\
    Home team &   0.430 &   (0.495) &   0.503 &   (0.500) &   0.578 &   (0.494) \\
                   Played on an unusual match day &   0.065 &   (0.247) &   0.074 &   (0.261) &   0.067 &   (0.250) \\
                   Played in the evening &   0.161 &   (0.368) &   0.158 &   (0.365) &   0.155 &   (0.362) \\
           Played after an international break &   0.124 &   (0.330) &   0.125 &   (0.330) &   0.134 &   (0.341) \\
          Played before an international break &   0.129 &   (0.335) &   0.128 &   (0.334) &   0.133 &   (0.339) \\
              WM/EM Season &   0.494 &   (0.500) &   0.500 &   (0.500) &   0.507 &   (0.500) \\
               Season after a WM/EM &   0.093 &   (0.291) &   0.083 &   (0.275) &   0.089 &   (0.284) \\
              Season before a WM/EM &   0.087 &   (0.282) &   0.091 &   (0.288) &   0.091 &   (0.287) \\
         Africa Cup Season &   0.510 &   (0.500) &   0.495 &   (0.500) &   0.495 &   (0.500) \\
         Match is played during Africa Cup &   0.152 &   (0.359) &   0.150 &   (0.357) &   0.142 &   (0.349) \\
Days since previous match (opponent) &   7.611 &   (6.127) &   7.490 &   (5.587) &   7.693 &   (5.733) \\
       Days to next match (opponent) &   7.234 &   (5.081) &   7.622 &   (5.742) &   7.555 &   (5.204) \\
         Upcoming Cup match (opponent) &   0.186 &   (0.389) &   0.152 &   (0.359) &   0.133 &   (0.340) \\
          Following Cup match (opponent) &   0.187 &   (0.390) &   0.156 &   (0.363) &   0.152 &   (0.359) \\
         Relegated last season (opponent) &   0.109 &   (0.311) &   0.123 &   (0.328) &   0.137 &   (0.344) \\
   Rank last season (opponent) &   8.941 &   (5.370) &   9.428 &   (5.170) &  10.211 &   (4.909) \\
       Number of previous clubs manager (opponent) &   2.598 &   (2.159) &   2.452 &   (2.032) &   2.309 &   (1.900) \\
    Number of league matches manager (opponent) & 130.438 & (148.815) & 126.196 & (139.250) & 116.838 & (132.946) \\
      Number of Top-5 matches manager (opponent)  & 166.936 & (194.943) & 152.520 & (171.142) & 133.470 & (151.504) \\
     Average market value quad ln (opponent) &  15.140 &   (0.832) &  15.031 &   (0.771) &  14.892 &   (0.702) \\
    Std. market value squad ln (opponent) &  15.091 &   (0.934) &  14.980 &   (0.879) &  14.838 &   (0.816) \\
        Average market value Top 11 ln (opponent) &  14.825 &   (0.879) &  14.710 &   (0.817) &  14.563 &   (0.747) \\
       Std. market value Top 11 ln (opponent) &  14.879 &   (0.974) &  14.790 &   (0.939) &  14.657 &   (0.897) \\   \\
	\midrule
	Observations   &  \multicolumn{2}{c}{20,721}          &  \multicolumn{2}{c}{19,953}         &  \multicolumn{2}{c}{18,294}\\  
	\midrule
        \end{tabular}
        \begin{tablenotes}
            \footnotesize
            \item \textbf{Notes:} Means and standard deviations (in parentheses) for all variables included in Table \ref{table_estimation_results_main}.
        \end{tablenotes}
        \label{table_desc_stat2}
    \end{threeparttable}
\end{table}

\begin{table}[H]
    \centering
    \begin{threeparttable}
    \caption{List of individual performance metrics} 
\begin{tabular}{lll}
\hline
 Variable                           & Variable                        & Variable                                   \\
\hline
 Playing time                       & accurate back zone pass         & accurate corners into box                   \\
accurate cross                     & accurate cross no corner         & accurate fwd zone pass                     \\
 accurate goal kicks                & accurate keeper throws          & accurate launches                          \\
 accurate layoffs                   & accurate long balls             & accurate pass                              \\
 accurate through ball              & accurate throws                 & aerial lost                                \\
 aerial won                         & att bx centre                   & att obx centre                             \\
 att bx right                       & att bx left                     & att corner                                 \\
 att fastbreak                      & att freekick goal               & att freekick target                        \\
 att freekick total                 & att freekick miss               & att freekick post                          \\
 att goal high centre               & att goal high left              & att goal high right                        \\
 att goal low centre                & att goal low left               & att goal low right                         \\
 att hd goal                        & att hd miss                     & att hd post                                \\
 att hd target                      & att hd total                    & att ibox blocked                           \\
 att ibox goal                      & att ibox miss                   & att ibox post                              \\
 att ibox target                    & att lf goal                     & att lf target                              \\
 att lf total                       & att miss high                   & att miss high left                         \\
 att miss high right                & att miss left                   & att miss right                             \\
 att one on one                     & att cmiss high                  & att cmiss high right                       \\
 att cmiss high left                & att cmiss left                  & att cmiss right                            \\
 att openplay                       & att pen goal                    & att pen miss                               \\
 att pen post                       & att pen target                  & att post high                              \\
 att post left                      & att post right                  & att setpiece                               \\
 attempts conceded ibox             & attempts conceded obox          & attempts ibox                              \\
 attempts obox                      & back pass                       & ball recovery                              \\
 blocked scoring att                & challenge lost                  & clean sheet                                \\
 clearance off-line                 & corner taken                    & cross not claimed                          \\
 crosses 18yardplus                 & dangerous play                  & dispossessed                               \\
 dive catch                         & dive save                       & duel lost                                  \\
 effective clearance                & effective head clearance        & error lead to goal                         \\
 error lead to shot                 & final third entries             & fouls                                      \\
 fouled final third                 & gk smother                      & goal assist                                \\
 goal assist intentional            & goal kicks                      & goals                                      \\
 goals conceded                     & goals conceded ibox             & goals conceded obox                        \\
 good high claim                    & goals open play                  & hand ball                                  \\
 head clearance                     & head pass                       & interceptions in box                       \\
 keeper pick up                     & keeper throws                   & last man tackle                            \\
 long pass own to opp               & long pass own to opp success    & offside provoked                           \\
 off target att assist               & on target att assist             & on target scoring att                       \\
 outfielder block                   & own goals                       & pen goals conceded                         \\
 penalty conceded                   & penalty save                    & penalty won                                \\
 post scoring att                   & punches                         & red card                                   \\
 saves                              & second yellow                   & second goal assist                         \\
 shot off target                    & total att assist                & total back zone pass                       \\
 total clearance                    & total contest                   & total corners into box                      \\
 total cross                        & total cross no corner            & total fastbreak                            \\
\midrule
        \end{tabular}
        \begin{tablenotes}
            \footnotesize
           \item \textbf{Notes:} This table includes a list of individual performance metrics used in Table \ref{table_estimation_results_main2}. The list is continued in Table \ref{table_metrics2}. fwd: forward; rf: right foot; lf: left foot; hd: header; att: attempt; ctrl: control; pen: penalty; ibox: in box.
        \end{tablenotes}
        \label{table_metrics1}
    \end{threeparttable}
\end{table}

\begin{table}[H]
    \centering
    \begin{threeparttable}
    \caption{List of individual performance metrics} 
\begin{tabular}{lll}
\hline
 Variable                           & Variable                        & Variable                                   \\
\hline
total fwd zone pass                & total high claim                & total launches                             \\
 total long balls                   & total offside                   & total pass                                 \\
 total scoring att                  & total sub off                   & total sub on                               \\
 total tackle                       & total through ball              & total throws                               \\
 touches                            & turnover                        & was fouled                                 \\
 won contest                        & won corners                     & won tackle                                 \\
 yellow card                        & penalty faced                   & goal assist open play                       \\
 goal assist setplay                & att assist open play             & att assist setplay                         \\
 overrun                            & interception won                & big chance created                         \\
 big chance missed                  & big chance scored               & unsuccessful touch                         \\
 fwd pass                           & backward pass                   & successful final third passes              \\
 total final third passes           & diving save                     & poss won def 3rd                           \\
 poss won mid 3rd                   & poss won att 3rd                & poss lost all                              \\
 poss lost ctrl                     & goal fastbreak                  & shot fastbreak                             \\
 pen area entries                   & hit woodwork                    & goal assist deadball                       \\
 freekick cross                     & accurate freekick cross         & open play pass                             \\
 successful open play pass          & attempted tackle foul           & fifty fifty                                \\
 successful fifty fifty             & blocked pass                    & failed to block                            \\
 put through                        & successful put through          & assist pass lost                           \\
 assist blocked shot                & assist attempt saved            & assist post                                \\
 assist free kick won               & assist handball won             & assist own goal                            \\
 assist penalty won                 & touches in opp box              & formation place                            \\
 times tackled                      & winning goal                    & interception                               \\
 expected goals                     & expected goals non penalty       & expected goals open play                    \\
 expected goals set play             & expected goals hd               & expected goals lf                          \\
 expected goals rf                  & expected goals freekick         & expected goals conceded                    \\
\midrule
        \end{tabular}
        \begin{tablenotes}
            \footnotesize
           \item \textbf{Notes:} This table includes a list of individual performance metrics used in Table \ref{table_estimation_results_main2}. The list is continued in Table \ref{table_metrics1}. fwd: forward; rf: right foot; lf: left foot; hd: header; att: attempt; ctrl: control; pen: penalty; ibox: in box.
        \end{tablenotes}
        \label{table_metrics2}
    \end{threeparttable}
\end{table}

\newpage
\section{Expected Goals}  \label{subsec: Appendix B}

In this section, we explain details on the calculation of expected goals. Expected goals (xG) quantify the quality of a scoring opportunity based on historical data from nearly a million shots. The name “expected goals” is derived from the mathematical concept of “expected value” and it is a measure of the likelihood of an outcome occurring. Variance from the expected value is inevitable, and provides valuable information, which we exploit in this paper.\footnote{https://theanalyst.com/eu/2023/08/what-is-expected-goals-xg/}

Each scoring opportunity is assigned a value on a scale from zero to one. A zero implies an impossible-to-score chance, while a one signifies a situation where a player is expected to score every time. This provides us with a way to express the likelihood of goals in different scenarios. For instance, if a chance inside the penalty area is assigned an xG of 0.1, it means that, on average, a player would score once in every ten attempts from that position, or 10\% of the time.


While watching a game, football enthusiasts often assess the likelihood of a goal being scored based on factors like the shooter's proximity to the goal, the shot angle, whether it's a one-on-one, or a header. However, individually evaluating the scoring likelihood for each of these unique scenarios would is a laborious task. xG models employ machine learning, specifically gradient boosting algorithms. The model leverages various variables, including those leading up to the shot and immediately before it. Over 20 factors are assessed to determine the likelihood of a goal. Some key factors include:

\begin{itemize}
	\item Distance to the goal
	\item Angle to the goal
	\item Goalkeeper's position, providing insight into their ability to make a save
	\item The shooter's view of the goal, considering the positions of other players
	\item The level of defensive pressure from opponents
	\item Shot type, including which foot the shooter used and whether it was a volley, header, or one-on-one
	\item Pattern of play (e.g., open play, fast break, direct free-kick, corner kick, throw-in, etc.)
	\item Information about the previous action, such as the type of assist (e.g., through ball, cross, etc.)
\end{itemize}

\begin{figure}[h!]

      \caption{Expected Goals by position}
    \begin{subfigure}{.3\textwidth}
      \centering
      \includegraphics[width=\textwidth]{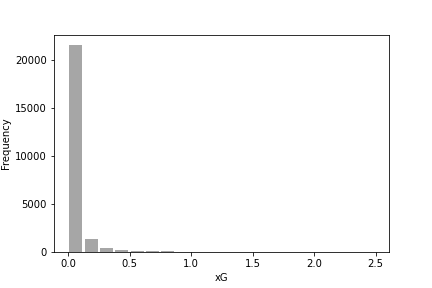}
      \caption{Defender}
    \end{subfigure}
    \hfill
        \begin{subfigure}{.3\textwidth}
      \centering
      \includegraphics[width=\textwidth]{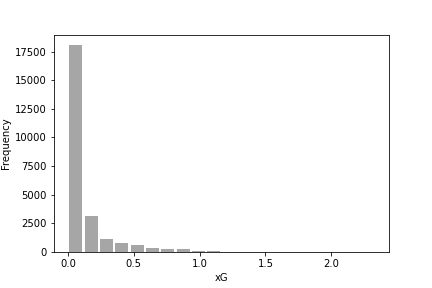}
      \caption{Midfielder}
    \end{subfigure}     
    \hfill
        \begin{subfigure}{.3\textwidth}
      \centering
      \includegraphics[width=\textwidth]{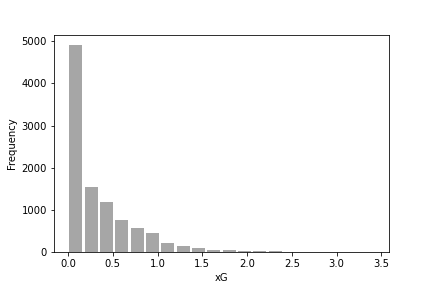}
      \caption{Striker}
    \end{subfigure}   
	\\ \footnotesize \textbf{Notes:} Distribution of expected goals players from different positions are exposed to during a match. For further discussion, see Section \ref{subsec: Measuring Shooting Performance}.\label{fig: expected goals_by_pos}
\end{figure}

We use expected goals as a measure of the quality of chances to separate shooting performance from collaborative efforts. Since the setting and the way of measuring individual performance are non-standard, we present further evidence that goals and xG allow for an evaluation of shooting performance. We conduct a test by correlating our shooting performance measure with expert-based assessments of shooting abilities.\footnote{For more details about the origin of these expert based skill assessments, see Section \ref{subsubsec: Control variables}.} Table \ref{tab:ability_corr} shows the beta coefficients of regressions of several specific ability measures on the number of goals scored conditional on xG and an overall assessment of players' strength. Reassuringly, we find that only the expert assessment of shooting ability is statistically significantly positively correlated with the number of goals scored. This result supports the argument that we are measuring shooting performance in a meaningful way and that shooting performance is an important aspect of the performance of soccer players.\par

\begin{table}[h!]
\caption{Player skills and shooting performance}
    \centering
    \begin{threeparttable}
        \begin{tabular}{lcccccc}
            \toprule
            & {$Shooting$} & {$Passing$} & {$Dribbling$} & {$Pace$} & {$Physical$} & {$Defense$} \\
            \midrule
            $\hat{\beta}$ & 0.346 & 0.164 & 0.207 & 0.123 & -0.266 & -0.815 \\
            $s.e.$ & [0.176] & [0.103] & [0.100] & [0.138] & [0.098] & [0.241] \\
            $p$-value & 0.013 & 0.110 & 0.126 & 0.371 & 0.007 & 0.000 \\
            \bottomrule
        \end{tabular}
        \begin{tablenotes}
            \small
            \item \textbf{Notes:} $\hat{\beta}$, standard error, and p-value of separate regressions of the form: $y_i = \beta goals_i + \gamma xG_i + \sigma X_i + \epsilon_i$. $Y_i$ is the ability measure based on the FIFA video game. $X_i$ is the overall quality of a player.
        \end{tablenotes}
    \end{threeparttable}
    \label{tab:ability_corr}
\end{table}

\newpage
\section{Further results and robustness checks}  \label{subsec: Appendix C}

\begin{figure}[h!]

      \caption{Common support figures}
    \begin{subfigure}{.3\textwidth}
      \centering
      \includegraphics[width=\textwidth]{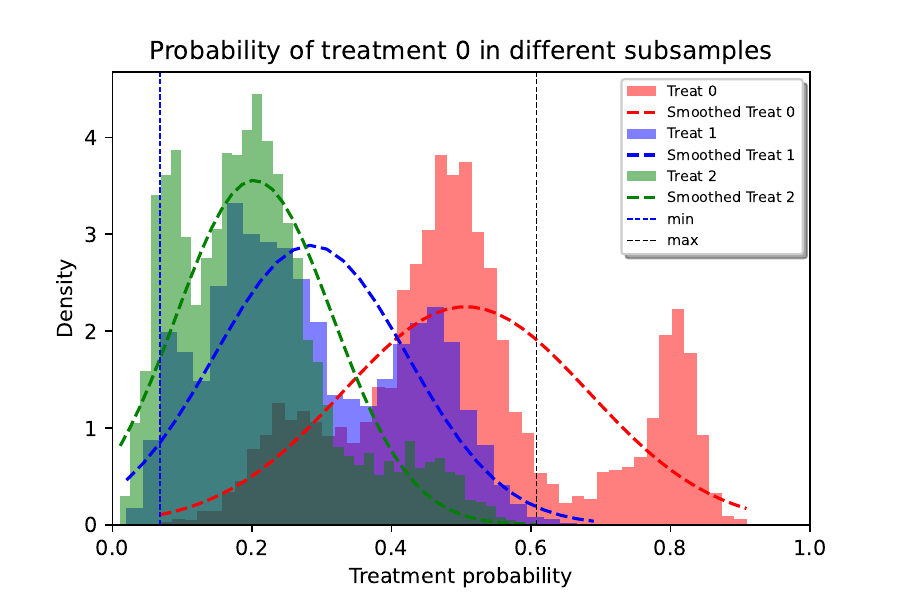}
      \caption{Treatment level: 0}
    \end{subfigure}
    \hfill
        \begin{subfigure}{.3\textwidth}
      \centering
      \includegraphics[width=\textwidth]{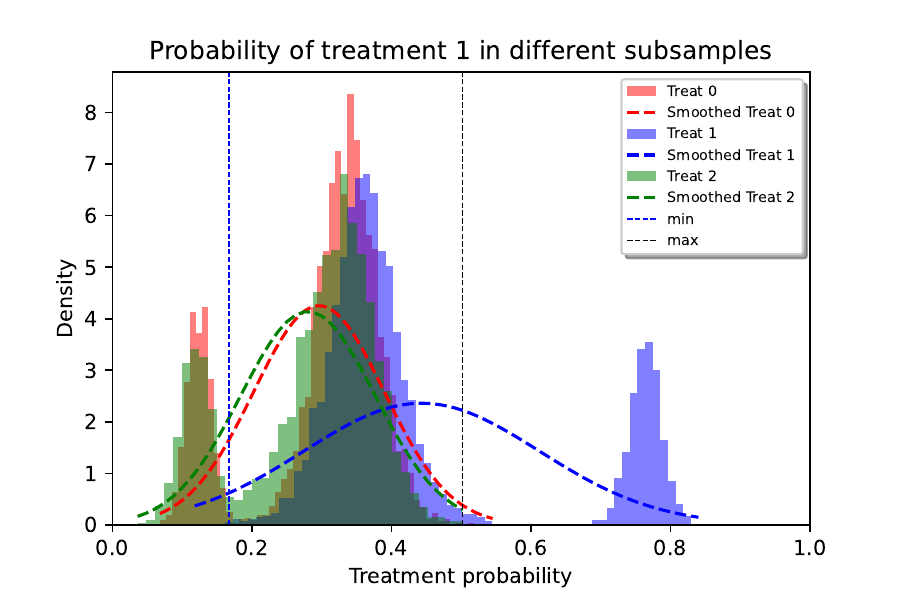}
      \caption{Treatment level: 1}
    \end{subfigure}     
    \hfill
        \begin{subfigure}{.3\textwidth}
      \centering
      \includegraphics[width=\textwidth]{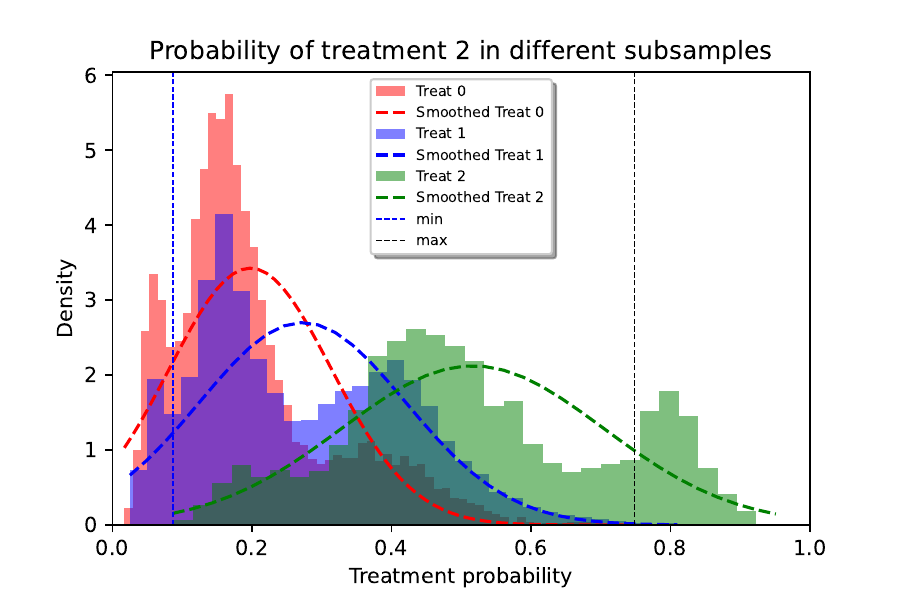}
      \caption{Treatment level: 2}
    \end{subfigure}   
	\\ \footnotesize \textbf{Notes:} Common support figures for each treatment level based on the specification in Column three of Table \ref{table_estimation_results_main} \label{fig: common_support}
\end{figure}

\clearpage

\begin{figure}[h!]

      \caption{Manager decisions - Distribution of estimated IATEs and overall effect heterogeneity}
    \begin{subfigure}{.49\textwidth}
      \centering
      \includegraphics[width=\textwidth]{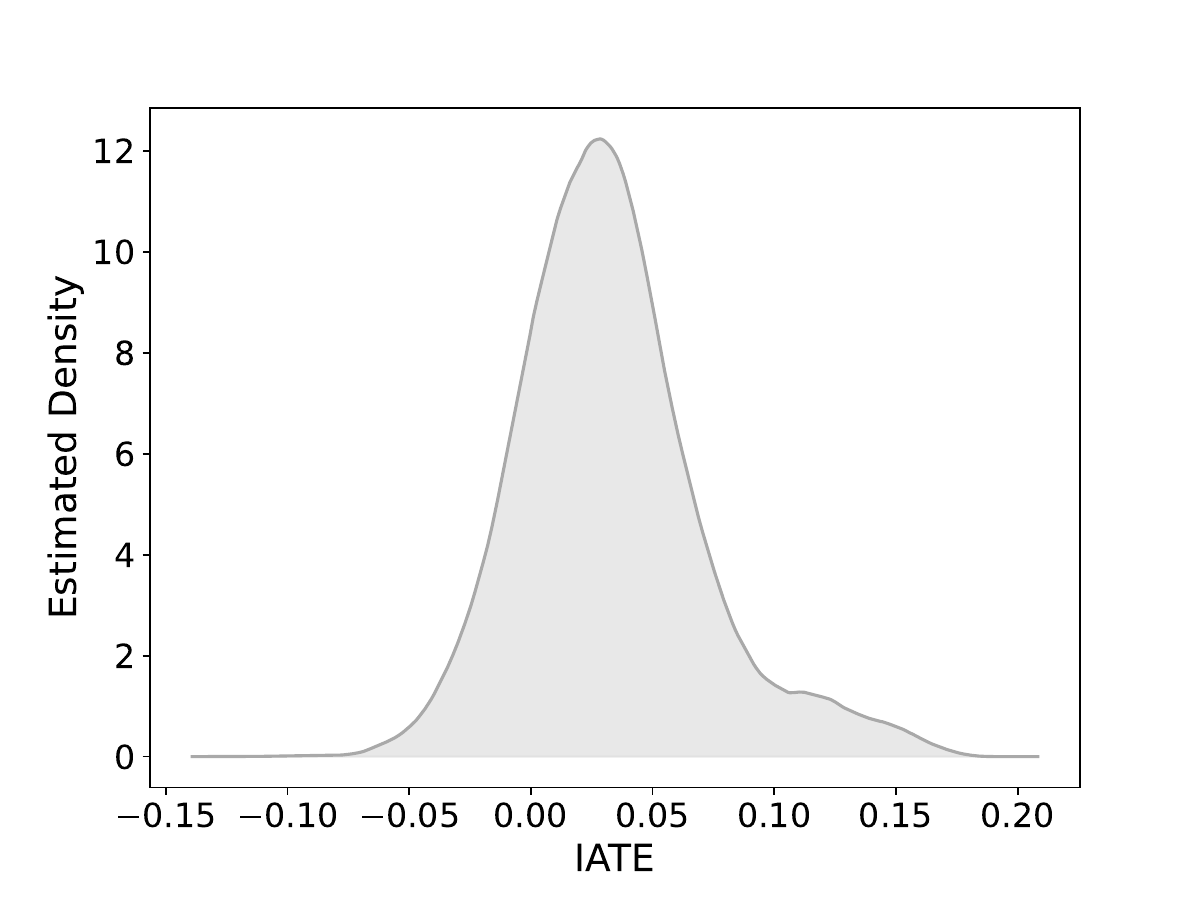}
      \caption{Manager}
    \end{subfigure}
    \hfill
        \begin{subfigure}{.49\textwidth}
      \centering
      \includegraphics[width=\textwidth]{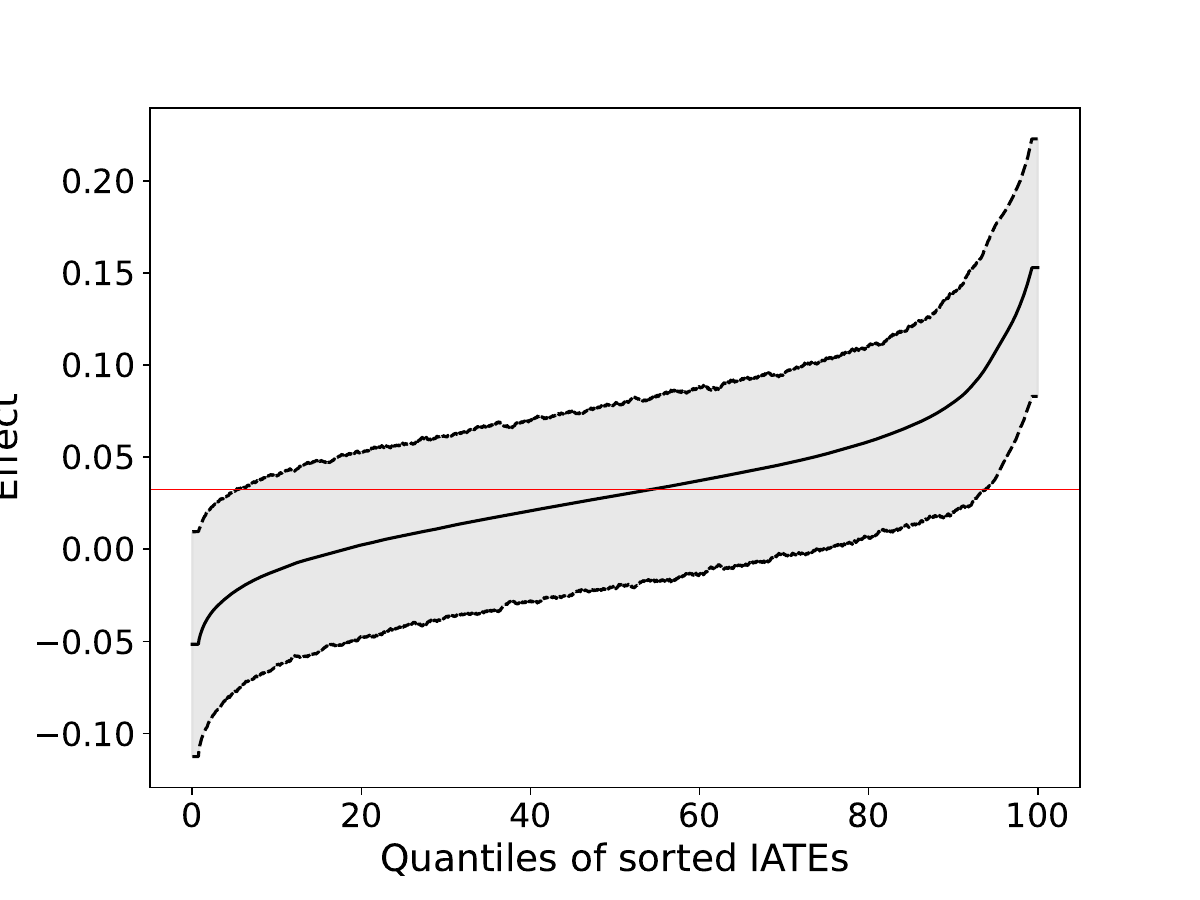}
      \caption{Manager}
    \end{subfigure}     
	\\ \footnotesize \textbf{Notes:} Estimates based on our main specification (Column 3 in Table \ref{table_estimation_results_main}). Panel a shows the distribution of estimated IATEs. The horizontal axis measures the effect of coworkers scoring two goals relative to coworkers scoring zero goals in the corresponding outcome variables. Kernel smooth
with Epanechnikov Kernel and Silverman (normality) bandwidth. Panel b shows IATEs sorted according to their size. 90\%-confidence interval of IATEs based on estimated standard errors and normal distribution. The vertical axis measures the average increase in performance evaluation from coworkers scoring two goals, relatively to coworkers scoring zero goals. The horizontal axis shows the rank of the ordered 
    observations. \label{fig_iate}
\end{figure}

\clearpage

\begin{table}[h!]
    \centering
    \begin{threeparttable}
        \caption{Robustness to treatment outliers}
        \begin{tabular}{lcccccc}
            \toprule
            Dependent Variable: & \multicolumn{3}{c}{Starting line-up next match} & \multicolumn{3}{c}{Expert evaluation} \\  
            \midrule
            2 vs. 0 & 0.036$^{***}$ & 0.040$^{***}$ & 0.037$^{***}$ & 0.410$^{***}$ & 0.365$^{***}$ & 0.315$^{***}$ \\   
            & (0.011) & (0.010) & (0.008) & (0.038) & (0.030) & (0.028) \\ 
            1 vs. 0 & 0.004 & 0.006 & 0.004 & 0.176$^{***}$ & 0.160$^{***}$ & 0.126$^{***}$ \\
            & (0.010) & (0.010) & (0.008) & (0.023) & (0.021) & (0.021) \\
            2 vs. 1 & 0.032$^{***}$ & 0.034$^{***}$ & 0.033$^{***}$ & 0.232$^{***}$ & 0.205$^{***}$ & 0.190$^{***}$\\ 
            & (0.011) & (0.010) & (0.010) & (0.036) & (0.028) & (0.026) \\
            \midrule
            Individual controls & $ \surd $ & $ \surd $ & $ \surd $ & $ \surd $ & $ \surd $ & $ \surd $\\ 
            Team controls & & $ \surd $ & $ \surd $ & & $ \surd $ & $ \surd $\\ 
            Match controls & & & $ \surd $ & & & $ \surd $\\ 
            Observations & 51,794 & 51,794 & 51,794 & 51,794 & 51,794 & 51,794 \\  
            \midrule
        \end{tabular}
        \begin{tablenotes}
            \footnotesize
            \item \textbf{Notes:} *, **, and *** represents statistical significance at the 10 \%, 5 \%, and 1 \% level respectively. Standard errors are clustered at the match level and presented in parentheses. All coefficients are population averages (ATE). For a full list of covariates included in each specification, see Table \ref{table_desc_stat2}.
        \end{tablenotes}
            \label{table_estimation_results_margin}
    \end{threeparttable}

\end{table}

\clearpage

\begin{table}[H]
    \centering
    \small
        \begin{threeparttable}
    \caption{OLS estimates}
\begin{tabular}{lcccc}
	\tabularnewline \midrule 
	Model:         & (1)             & (2)             & (3) &  (4)\\  
	\midrule
	\emph{Starting line-up next match}\\
	Coworker performance         & 0.023$^{***}$ & 0.020$^{***}$ & 0.020$^{***}$ & 0.019$^{***}$\\   
	&    (0.003)     &    (0.003)     & (0.002) & (0.002)\\  
	\emph{Expert evaluation}\\
	Coworker performance        & 0.234$^{***}$ & 0.234$^{***}$ & 0.227$^{***}$ & 0.224$^{***}$\\   
	&   (0.011)      &   (0.010)      &  (0.010) & (0.010)\\
	\midrule
    Individual controls & $ \surd $  &        &         & \\
    Team controls & $ \surd $  &           & $ \surd $          & \\
    Match controls & $ \surd $  &           & $ \surd $          & $ \surd $\\
	Individual FE &   & $ \surd $          & $ \surd $          & \\
	Individual-Season FE    &          &     &      & $ \surd $ \\ 
	Observations   & 58,968          & 58,968          & 58,968 & 58,968 \\  
\midrule
        \end{tabular}
        \begin{tablenotes}
            \footnotesize
            \item \textbf{Notes:} Linear regression. *, **, and *** represents statistical significance at the 10 \%, 5 \%, and 1 \%, respectively. We additionally control for the quality of chances coworkers are exposed to (xG). 
        \end{tablenotes}
        \label{table_femodel}
    \end{threeparttable}
\end{table}

\clearpage

\begin{table}[h!]
    \caption{Assignment rules of shallow trees}
    \centering
    \small
    \begin{threeparttable}
        \begin{tabularx}{\linewidth}{l *{2}{X}}
            \toprule
            \multicolumn{3}{c}{Tree depth = 3}\\  
            \midrule
            0 Goals & 1 Goal & 2 Goals \\
            \midrule
            Managers with more than 258 matches      \\
            in Top-5 leagues \& \\
            Player quality above 75 \&  \\
            Team strength above 74.654  \\
            or & Nobody     & All other\\
            Managers with less than 259 matches  \\
            in Top-5 leagues \& \\
            Player quality above 83 \& \\
            Team strength above 74.654 \\
            \midrule    
        \end{tabularx}
        \begin{tablenotes}
            \footnotesize
            \item \textbf{Notes:} The outcome variable used is a binary indicator, whether a player is fielded in the starting line-up in the next match. For the sake of brevity, we do not list the specific sectors selected by the decision tree.
        \end{tablenotes}
    \end{threeparttable}
    \label{table_policytree_Startelf}
\end{table}

\clearpage

\textbf{Results from k-means clustering.} K-means clustering groups similar data points into distinct clusters by assigning each point to the cluster whose mean is closest. The "k" in k-means denotes the predetermined number of clusters. The algorithm iteratively adjusts centroids to minimize within-cluster variance, resulting in a partitioning of data into k clusters. In our implementation, k-means$++$ clustering utilizes Individualized Average Treatment Effects (IATEs) relative to the no-goal treatment to categorize players into clusters.\footnote{K-means$++$ compared to k-means clustering improves initial center selection for faster computation.}

\begin{table}[H]
    \centering
    \small
    \begin{threeparttable}
    \caption{Means of IATEs and covariates by cluster obtained from k-means clustering} 
\begin{tabular}{lcccc}
		\tabularnewline \midrule 
		\multicolumn{1}{c}{Cluster}& \multicolumn{1}{c}{1}& \multicolumn{1}{c}{2}& \multicolumn{1}{c}{3}& \multicolumn{1}{c}{4}\\  
		\midrule
         & \multicolumn{4}{c}{IATE for the comparison 2 vs 0} \\
         \midrule
        & -0.010     & 0.025    & 0.059 & 0.116 \\
		\midrule
         & \multicolumn{4}{c}{Mean of selected features}\\
         \midrule
        xG Coworker &  1.31   &   1.17    &  1.23 & 0.23\\ 
		Player quality (Fifa) &   76.79  &   74.89    &  73.52 & 73.53 \\ 
        Defensive Skills (Fifa) & 62.77 & 60.53 & 56.06 & 55.70 \\
        Pace (Fifa) &  71.86 & 71.08  & 71.91 & 71.22 \\
        Passing Skills (Fifa) & 66.74  & 66.25  & 66.80  & 64.79 \\
        Physical Skills (Fifa) & 72.46 &  71.28 &  69.46 &  70.20\\
        Dribbling Skills (Fifa) & 69.41 & 69.10 & 70.52 & 68.60 \\
        Shooting Skills (Fifa) & 58.93  &  58.79 &  60.86 & 59.55 \\
        Time in club &   3.03  &  2.45      & 1.89 & 2.00 \\ 
        Age & 26.08 & 26.16 & 26.07 & 25.91\\
        German &   0.44  &  0.44 & 0.42      &  0.43\\ 
        Position &  3.65   &   3.73     &  3.78 & 3.75\\ 
        Manager experience &  174.15   & 144.92       & 140.32 & 143.41 \\ 
        Team strength (Fifa) &  72.40   &   71.02     & 70.24 & 70.29 \\
        Relegated from 2nd to 1st division & 0.05 & 0.11 & 0.20 & 0.19\\
        Rank last season & 6.83 & 9.62 & 11.50 & 11.35 \\
        Opponent strength (MW) &  15.33   &   14.99     &  14.66 & 15.22\\ 
        Rank last season opponent & 7.98  &  9.73 & 11.21  & 8.49 \\        
\midrule
        \end{tabular}
        \begin{tablenotes}
            \footnotesize
            \item \textbf{Notes:} Summary of results from k-means clustering approach. Cluster 1 is the least beneficial group, while cluster 4 is the most beneficial group. In the bottom part, we drop a set of covariates for which we do not observe meaningful differences across cluster. The outcome variable used is a binary indicator, whether a player is fielded in the starting line-up in the next match.
        \end{tablenotes}
        \label{table_cluster_Startelf}
    \end{threeparttable}
\end{table}

The analysis reveals substantial heterogeneity in the degree of spillover effects. The spillover effects on fielding decisions in the next match range from 0.01 to 0.12. To facilitate interpretation, let's consider one examples. The average chances of being replaced in the match are around 20\%. In the group most affected by spillover effects, there is a more than 50\% decrease in the likelihood of being replaced in the next match. 

In the bottom part of Table \ref{table_cluster_Startelf} we report means per cluster for selected features. For manager decisions, we find that the group benefiting the least from spillover effects is broadly characterized by higher team performance (xG) and higher individual quality—especially marked by superior defensive skills. On average, this group has also spent more time in the club, and the club manager is more experienced compared to the most beneficial group. As observed in previous sections, team strength and position do not exert a substantial impact.

\clearpage

\begin{table}[H]
    \centering
    \begin{threeparttable}
        \caption{Own shooting performance and performance evaluation}
        \begin{tabular}{lccc}
            \toprule
            & (2 vs 0) & (1 vs 0) & (2 vs 1) \\  
            \midrule
            \emph{Starting line-up next match}\\
            Coworker performance & 0.100$^{**}$ & 0.086$^{***}$ & 0.014\\   
            & (0.044) & (0.011) & (0.045) \\ 
            \emph{Expert evaluation}\\
            Coworker performance & 1.965$^{***}$ & 1.058$^{***}$ & 0.907$^{***}$\\   
            & (0.256) & (0.028) & (0.257) \\
            \midrule
            Individual controls & $ \surd $ & $ \surd $ & $ \surd $\\ 
            Team controls & $ \surd $ & $ \surd $ & $ \surd $\\ 
            Match controls & $ \surd $ & $ \surd $ & $ \surd $ \\ 
            Observations & 58,968 & 58,968 & 58,968\\  
            \midrule
        \end{tabular}
        \begin{tablenotes}
            \footnotesize
            \item \textbf{Notes:} *, **, and *** represents statistical significance at the 10\%, 5\%, and 1\% level respectively. Standard errors are clustered at the match level and sre presented in parentheses. All effects are population averages (ATE). For a full list of covariates included in each specification, see Table \ref{table_desc_stat2}.
        \end{tablenotes}
        \label{table_estimation_results_own}
    \end{threeparttable}
\end{table}

\clearpage

\begin{table}[H]
    \centering
    \small
    \begin{threeparttable}
    \caption{Subsample analysis - Type of coworker and type of worker}
	\begin{tabular}{lccc}
		\midrule
		Dependent Variable: & \multicolumn{3}{c}{Manager decision}\\
		& (All)      & (Defender)                       & (Striker)  \\  
		\midrule
		\emph{Defender}\\
		Shooting Performance   & 0.019$^{*}$      & 0.021$^{*}$  & 0.013\\   
		& (0.011) &   (0.012)       & (0.011)\\  
		\emph{Striker}\\
		Shooting Performance   & 0.011     & 0.011  & 0.004\\   
		&  (0.010) &  (0.010)         & (0.016)\\  
		\midrule
		Individual controls & $ \surd $  & $ \surd $             & $ \surd $\\ 
		Team controls & $ \surd $  &    $ \surd $           & $ \surd $\\ 
		Match controls & $ \surd $  &    $ \surd $         & $ \surd $ \\ 
		Observations & 58,968  & 23,293             & 10,044\\ \midrule
        \end{tabular}
        \begin{tablenotes}
            \footnotesize
            \item \textbf{Notes:} Subsample analysis based on types of coworkers. The outcome variable is the manager decision. *, **, and *** represents statistical significance at the 10 \%, 5 \%, and 1 \%, respectively. 
        \end{tablenotes}
        \label{table_estimation_results_main3}
    \end{threeparttable}
\end{table}

\clearpage

\begin{table}[H]
    \centering
    \small
    \begin{threeparttable}
    \caption{Subsample analysis - Type of coworker and type of worker}
	\begin{tabular}{lccc}
		\midrule
		Dependent Variable: & \multicolumn{3}{c}{Expert evaluation}\\
		& (All)      & (Defender)                     & (Striker)  \\  
		\midrule
		\emph{Defender}\\
		Shooting Performance   & 0.080$^{***}$      & 0.057$^{**}$ & 0.101$^{***}$\\   
		& (0.024) &   (0.026)      & (0.028)\\  
		\emph{Striker}\\
		Shooting Performance   & 0.137$^{***}$      & 0.127$^{***}$  & 0.146$^{***}$\\   
		&  (0.020) &  (0.019)       & (0.031)\\  
		\midrule
		Individual controls & $ \surd $  & $ \surd $                  & $ \surd $\\ 
		Team controls & $ \surd $  &    $ \surd $          & $ \surd $\\ 
		Match controls & $ \surd $  &    $ \surd $          & $ \surd $ \\ 
		Observations & 58,968  & 23,293                 & 10,044\\ \midrule
        \end{tabular}
        \begin{tablenotes}
            \footnotesize
            \item \textbf{Notes:} Subsample analysis based on types of coworkers. The outcome variable is the journalists' ratings. *, **, and *** represents statistical significance at the 10 \%, 5 \%, and 1 \%, respectively. 
        \end{tablenotes}
        \label{table_estimation_results_main4}
    \end{threeparttable}
\end{table}

\newpage
\section{Effect heterogeneity journalists' ratings}  \label{subsec: Appendix D}

\begin{figure}[h!]

      \caption{GATEs for different player characteristics}
    \begin{subfigure}{.49\textwidth}
      \includegraphics[width=\textwidth]{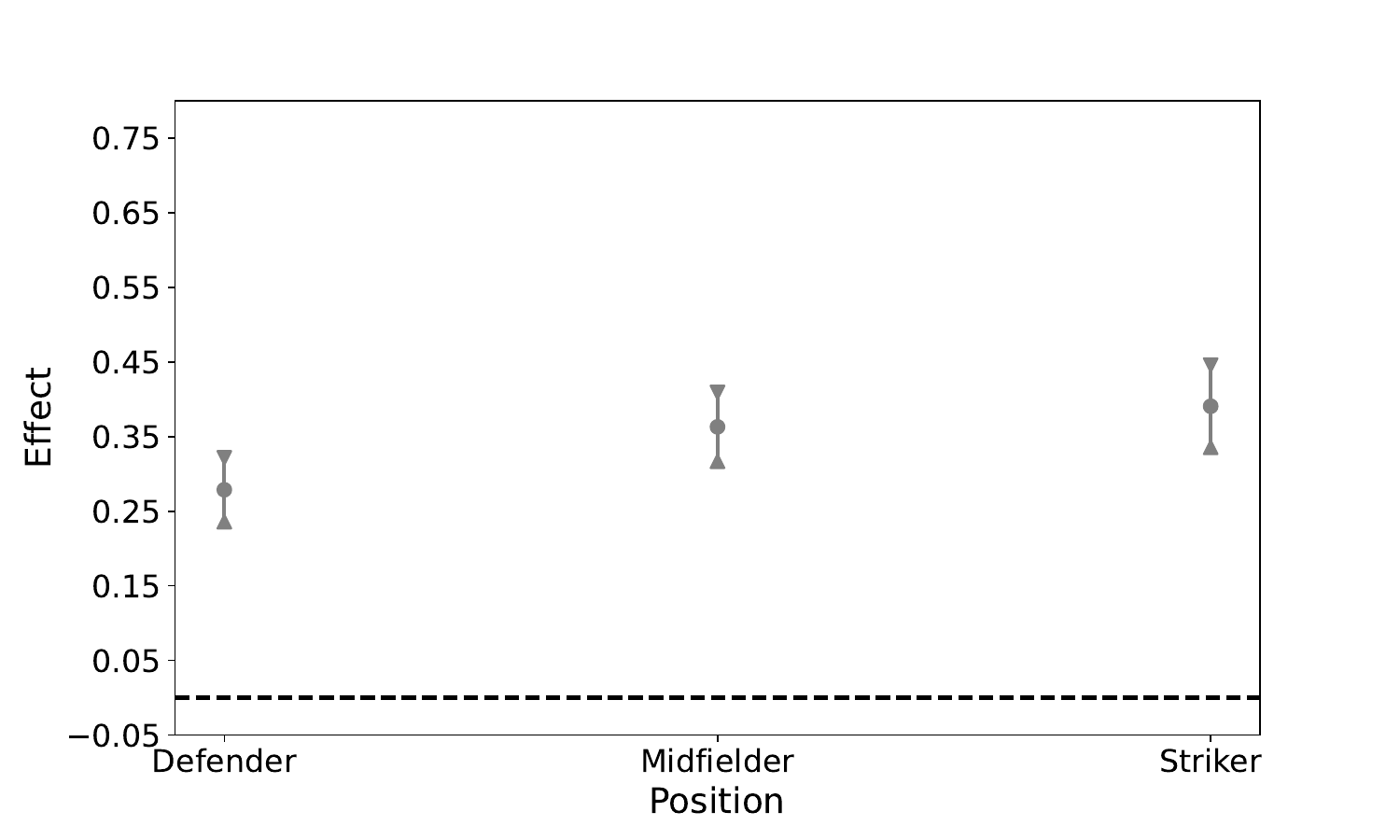}
      \caption{Expert}
    \end{subfigure}
    \hfill
        \begin{subfigure}{.49\textwidth}
      \includegraphics[width=\textwidth]{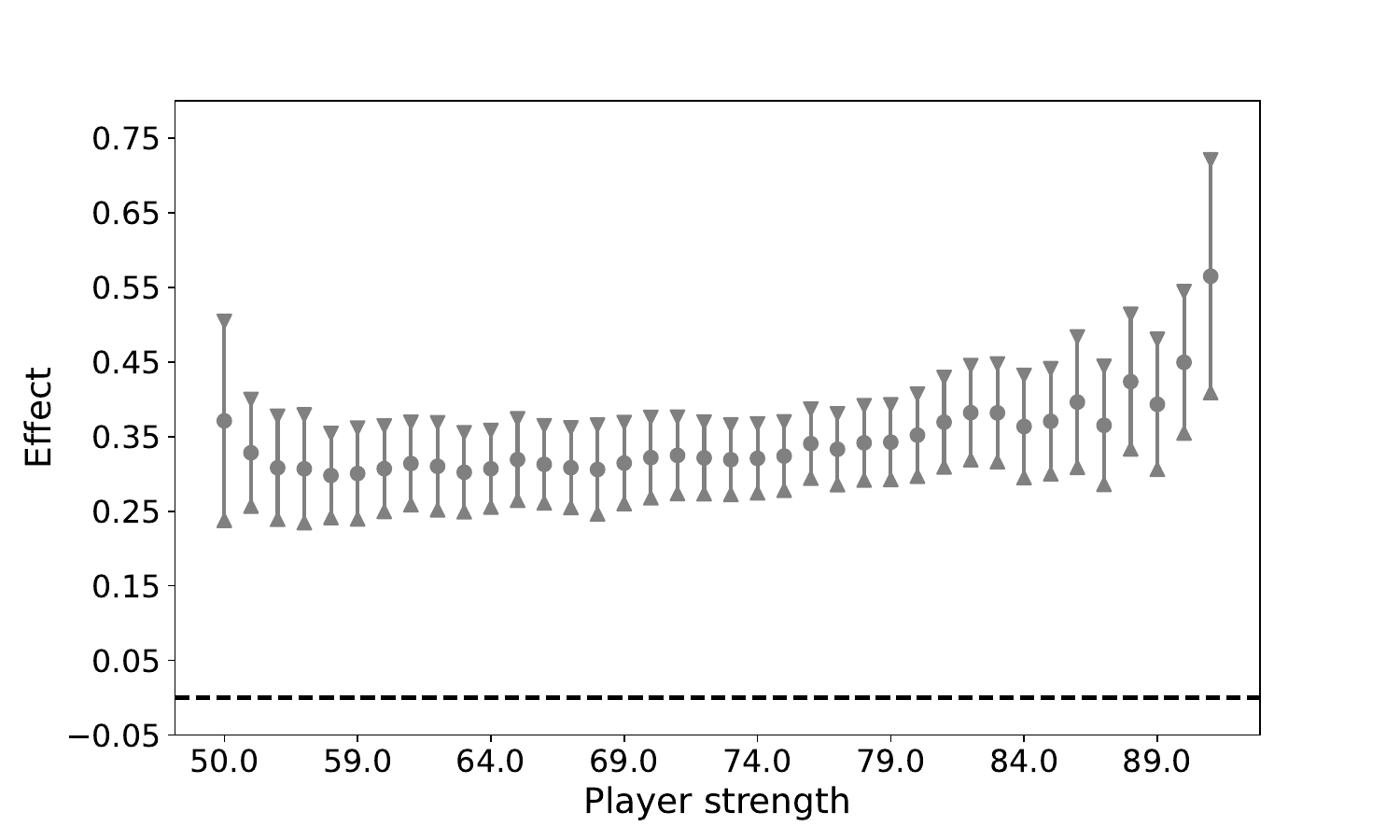}
      \caption{Expert}
    \end{subfigure}   
    \hfill
    \begin{center}
        \begin{subfigure}{.49\textwidth}
      \includegraphics[width=\textwidth]{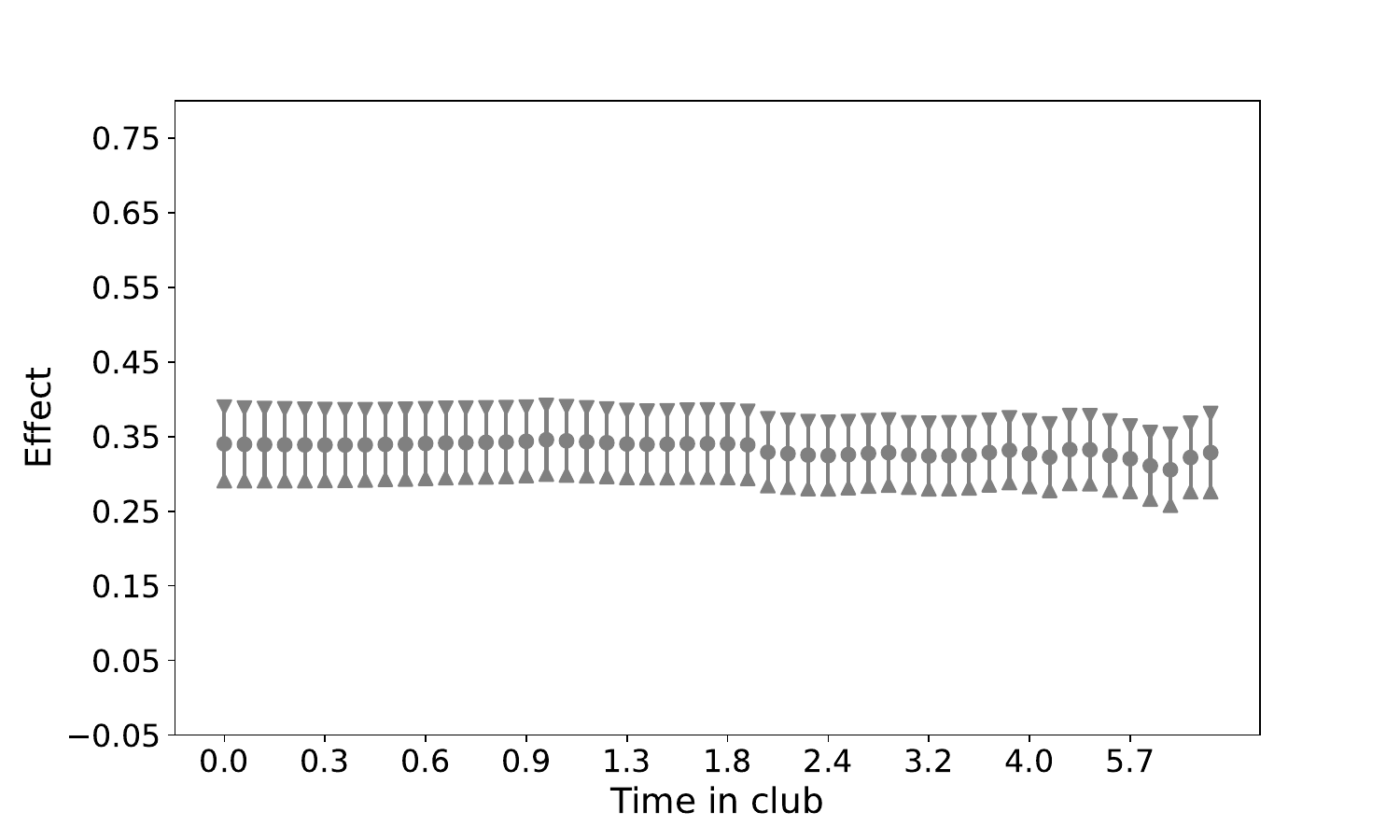}
      \caption{Expert}
    \end{subfigure}              
    \end{center}
	\footnotesize \textbf{Notes:} The vertical axis denotes the respective GATE and its 90\% confidence interval for the comparison of coworkers scoring two goals, relatively to coworkers scoring zero goals. \label{fig_cates2}
\end{figure}

\clearpage

\begin{figure}[h!]

      \caption{Journalists' ratings - Distribution of estimated IATEs and overall effect heterogeneity}
    \begin{subfigure}{.49\textwidth}
      \centering
      \includegraphics[width=\textwidth]{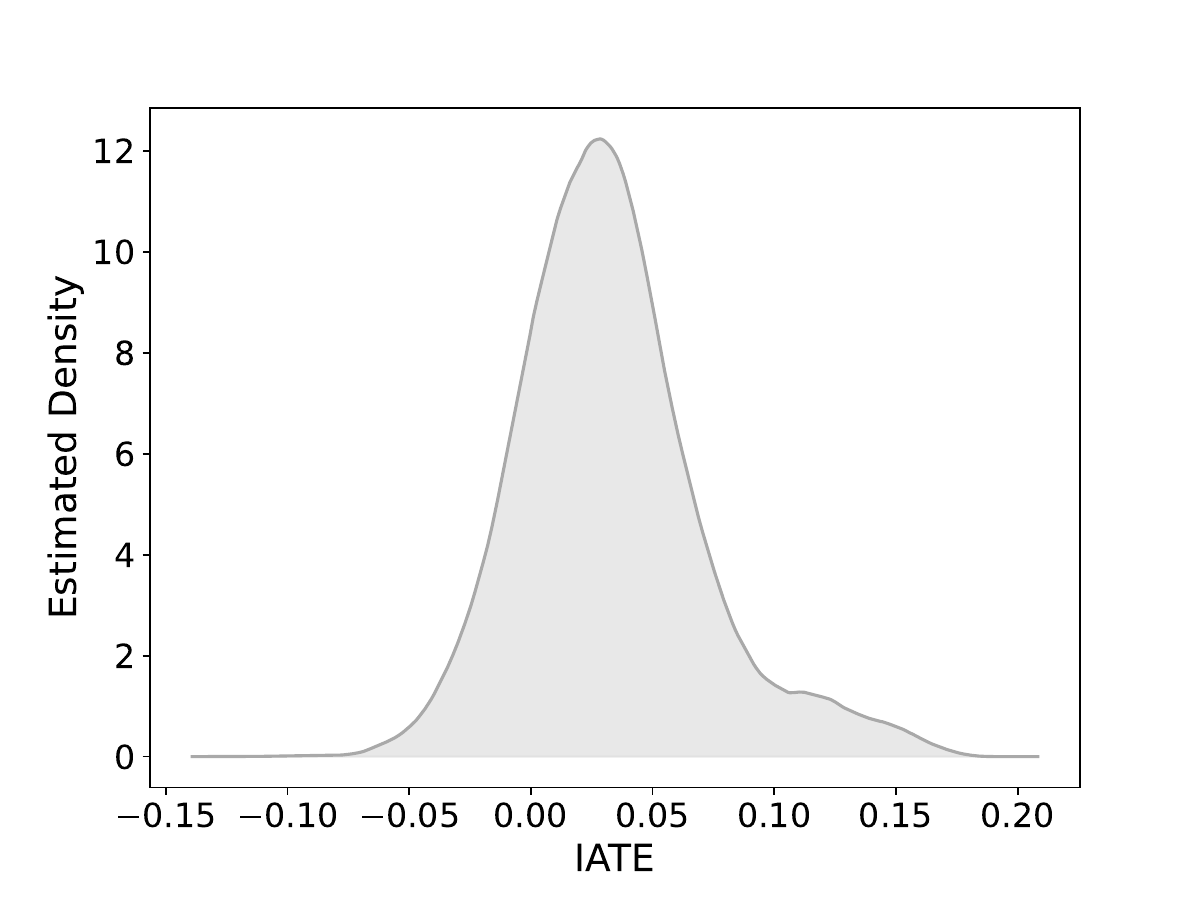}
      \caption{Manager}
    \end{subfigure}
    \hfill
        \begin{subfigure}{.49\textwidth}
      \centering
      \includegraphics[width=\textwidth]{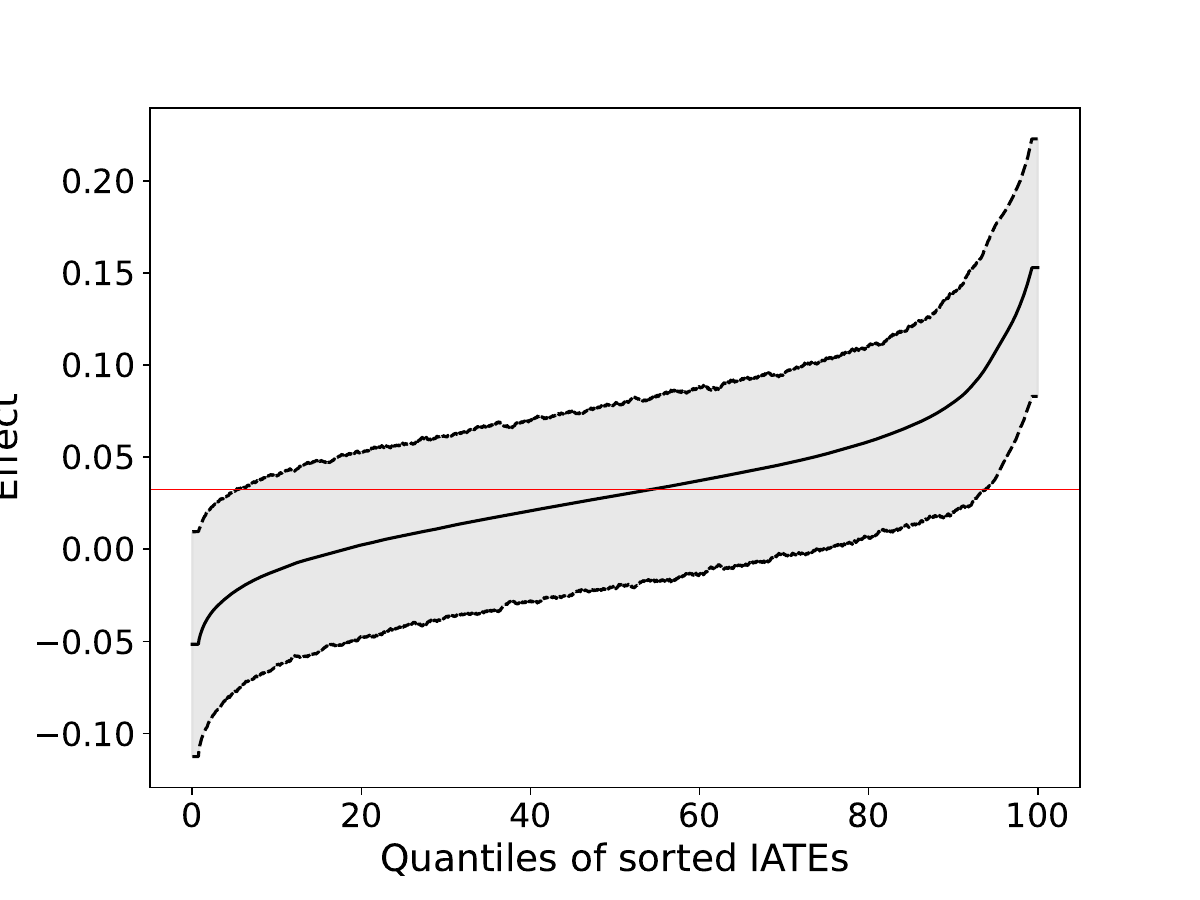}
      \caption{Manager}
    \end{subfigure}     
	\\ \footnotesize \textbf{Notes:} Estimates based on our main specification (Column 3 in Table \ref{table_estimation_results_main}). Panel a shows the distribution of estimated IATEs. The horizontal axis measures the effect of coworkers scoring two goals relative to coworkers scoring zero goals in the corresponding outcome variables. Kernel smooth
with Epanechnikov Kernel and Silverman (normality) bandwidth. Panel b shows IATEs sorted according to their size. 90\%-confidence interval of IATEs based on estimated standard errors and normal distribution. The vertical axis measures the average increase in performance evaluation from coworkers scoring two goals, relatively to coworkers scoring zero goals. The horizontal axis shows the rank of the ordered 
    observations. \label{fig_iate2}
\end{figure}

\clearpage

\textbf{Results k-means clustering.} Since we could not derive informative results for the journalists' ratings using the policy simulation exercise, as all observation were assigned to the highest treatment level, we depict the dependence of the effects on covariates using k-means++ clustering \cite{arthur2007k}. The analysis reveals substantial heterogeneity in the degree of spillover effects. The spillover effects on journalists' evaluations range from 0.18 in the least affected cluster to 0.51 in the most affected cluster. To facilitate interpretation, let's consider one examples. A 0.51 increase in the evaluation received by journalists is roughly equivalent to 0.5 times the effect of scoring one goal oneself, given the faced chances. The size of the spillover effect is, thus, one-quarter of the effect of one's own performance in the same task.\footnote{Results used for this comparison are shown in Appendix C Table \ref{table_estimation_results_own}.}

In the bottom part of Table \ref{table_cluster_Expert}, we report means per cluster for selected features. The group benefiting the most from spillover effects is roughly characterized by higher team performance (xG), with individual quality negatively correlated with the treatment effect size. Specifically, players with high defensive skills are prevalent in the group of least beneficial players, while players with high offensive skills are in the group that benefits the most from coworker shooting performance. These findings align with the fact that strikers are more likely to be in the group of most beneficial players. Manager and team strength matter less compared to the manager's decision.

\begin{table}[H]
    \centering
    \small
    \begin{threeparttable}
    \caption{Means of IATEs and covariates by cluster obtained from k-means clustering.}
	\begin{tabular}{lcccccc}
		\midrule
		\multicolumn{1}{c}{Cluster}& \multicolumn{1}{c}{1}& \multicolumn{1}{c}{2}& \multicolumn{1}{c}{3}&  \multicolumn{1}{c}{4} & \multicolumn{1}{c}{5} &\multicolumn{1}{c}{6}\\  
		\midrule
         & \multicolumn{6}{c}{IATE for the comparision 2 vs 0} \\
         \midrule
        & 0.179     & 0.250    & 0.290 & 0.371 & 0.387 & 0.510 \\
		\midrule
         & \multicolumn{6}{c}{Mean of selected features}\\
         \midrule
        xG Coworker &  2.08   &   1.37    &  0.75 & 1.09 & 0.91 & 1.51\\ 
		Player quality (Fifa) &   75.98  &   74.31    &  73.78 & 74.73 & 74.72 & 78.29 \\ 
        Defensive Skills (Fifa) & 74.83 & 67.61 & 63.46 & 57.52 & 53.01 & 42.11 \\
        Pace (Fifa) &  67.02 & 69.46  & 70.08 & 71.53 & 73.30 & 77.93 \\
        Passing Skills (Fifa) & 61.88  & 63.85  & 64.70  & 67.37 & 68.23 & 72.44 \\
        Physical Skills (Fifa) & 75.17 &  72.86 &  71.78 &  70.76 & 69.31 & 66.83\\
        Dribbling Skills (Fifa) & 61.66 & 65.49 & 66.99 & 70.91 & 72.61 & 79.08 \\
        Shooting Skills (Fifa) & 47.74  &  52.79 &  55.70 & 62.24 & 64.17 & 73.10 \\
        Time in club &   3.12  &  2.60      & 2.32 & 2.47 & 2.17 & 2.15 \\ 
        Age & 26.29 & 26.35 & 26.17 & 26.20 & 25.69 & 25.86\\
        German &   0.45  &  0.46 & 0.46      &  0.43 & 0.42 & 0.37\\ 
        Position &  2.20 &   2.52     &  2.72 & 3.13 & 3.23 & 3.72\\ 
        Manager experience &  148.88   & 142.55  & 104.94 & 191.98 & 133.05 & 216.83 \\ 
        Team strength (Fifa) &  71.94   &   70.83     & 70.37 & 71.19 & 71.02 & 72.56 \\
        Relegated from 2nd to 1st division & 0.07 & 0.15 & 0.17 & 0.13 & 0.10 & 0.06\\
        Rank last season & 7.21 & 9.95 & 10.75 & 9.66 & 9.59 & 7.54 \\
        Opponent strength (MW) &  15.01   &   14.98 & 15.06 & 15.06 & 14.94 & 14.93 \\ 
        Rank last season opponent & 9.68  &  9.72 & 9.23  & 9.35 & 9.82 & 10.06 \\ 
        \midrule
        \end{tabular}
        \begin{tablenotes}
            \footnotesize
            \item \textbf{Notes:} Summary of results from k-means clustering approach. Cluster 1 is the least beneficial group, while cluster 6 is the most beneficial group. In the bottom part, we drop a set of covariates for which we do not observe meaningful differences across cluster. The outcome variable are the journalists’ evaluations.
        \end{tablenotes}
        \label{table_cluster_Expert}
    \end{threeparttable}
\end{table}

\end{appendices}
\end{document}